\documentclass{elsarticle}

\usepackage[reviewcopy]{adndt}
\usepackage{longtable}

\usepackage{amsmath}

\usepackage{amssymb}

\biboptions{square,sort&compress}
\bibpunct[]{[}{]}{,}{n}{}{;}
\begin{document}

\begin{frontmatter}

\journal{Atomic Data and Nuclear Data Tables}


\title{Discovery of Tantalum, Rhenium, Osmium, and Iridium Isotopes}

\author{R. Robinson}
\author{M. Thoennessen\corref{cor1}}\ead{thoennessen@nscl.msu.edu}

 \cortext[cor1]{Corresponding author.}

 \address{National Superconducting Cyclotron Laboratory and \\ Department of Physics and Astronomy, Michigan State University, \\ East Lansing, MI 48824, USA}

\begin{abstract}
Currently, thirty-eight tantalum, thirty-eight rhenium, thirty-nine osmium, and thirty-eight iridium, isotopes have been observed and the discovery of these isotopes is discussed here. For each isotope a brief synopsis of the first refereed publication, including the production and identification method, is presented.
\end{abstract}

\end{frontmatter}





\newpage
\tableofcontents
\listofDtables

\vskip5pc

\section{Introduction}\label{s:intro}

The discovery of tantalum, rhenium, osmium, and iridium isotopes is discussed as part of the series summarizing the discovery of isotopes, beginning with the cerium isotopes in 2009 \cite{2009Gin01}. Guidelines for assigning credit for discovery are (1) clear identification, either through decay-curves and relationships to other known isotopes, particle or $\gamma$-ray spectra, or unique mass and Z-identification, and (2) publication of the discovery in a refereed journal. The authors and year of the first publication, the laboratory where the isotopes were produced as well as the production and identification methods are discussed. When appropriate, references to conference proceedings, internal reports, and theses are included. When a discovery includes a half-life measurement the measured value is compared to the currently adopted value taken from the NUBASE evaluation \cite{2003Aud01} which is based on the ENSDF database \cite{2008ENS01}. In cases where the reported half-life differed significantly from the adopted half-life (up to approximately a factor of two), we searched the subsequent literature for indications that the measurement was erroneous. If that was not the case we credited the authors with the discovery in spite of the inaccurate half-life. All reported half-lives inconsistent with the presently adopted half-life for the ground state were compared to isomers half-lives and accepted as discoveries if appropriate following the criterium described above.


Good examples why publications in conference proceedings should not be considered are $^{118}$Tc and $^{120}$Ru which had been reported as being discovered in a conference proceeding \cite{1996Cza01} but not in the subsequent refereed publication \cite{1997Ber01}.

The initial literature search was performed using the databases ENSDF \cite{2008ENS01} and NSR \cite{2008NSR01} of the National Nuclear Data Center at Brookhaven National Laboratory. These databases are complete and reliable back to the early 1960's. For earlier references, several editions of the Table of Isotopes were used \cite{1940Liv01,1944Sea01,1948Sea01,1953Hol02,1958Str01,1967Led01}. A good reference for the discovery of the stable isotopes was the second edition of Aston's book ``Mass Spectra and Isotopes'' \cite{1942Ast01}.

\section{Discovery of $^{155-192}$Ta}

Thirty-eight tantalum isotopes from A = 155--192 have been discovered so far; these include 1 stable, 26 neutron-deficient and 11 neutron-rich isotopes. According to the HFB-14 model \cite{2007Gor01}, $^{246}$Ta should be the last odd-odd particle stable neutron-rich nucleus while the odd-even particle stable neutron-rich nuclei should continue through $^{249}$Ta. The proton dripline has already been crossed with the observation of the proton emitters $^{155}$Ta and $^{156}$Ta, however, about seven additional proton-rich tantalum isotopes could still have half-lives longer than 10$^{-9}$~s \cite{2004Tho01}. Thus, about 63 isotopes have yet to be discovered corresponding to 63\% of all possible yttrium isotopes.

\begin{figure}
	\centering
	\includegraphics[scale=.7]{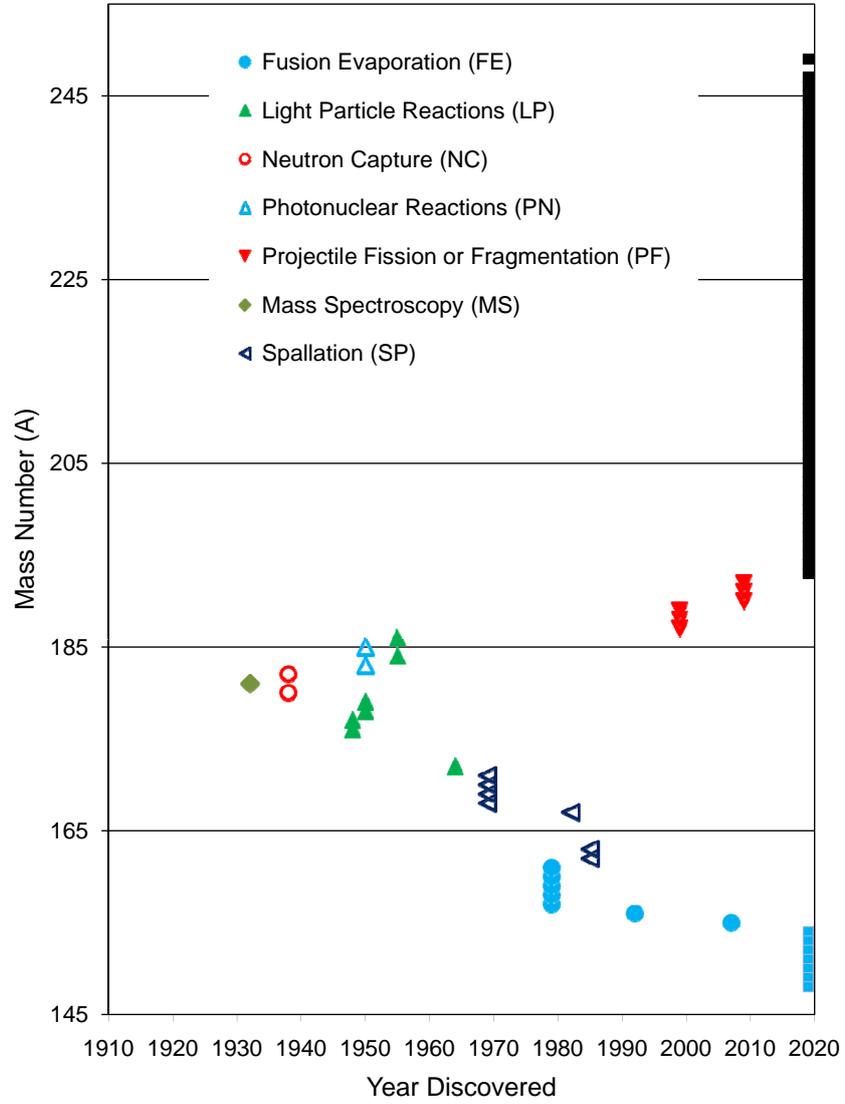}
	\caption{Tantalum isotopes as a function of time when they were discovered. The different production methods are indicated. The solid black squares on the right hand side of the plot are isotopes predicted to be bound by the HFB-14 model. On the proton-rich side the light blue squares correspond to unbound isotopes predicted to have half-lives larger than $\sim 10^{-9}$~s.}
\label{f:year-ta}
\end{figure}

Figure \ref{f:year-ta} summarizes the year of first discovery for all tantalum isotopes identified by the method of discovery. The range of isotopes predicted to exist is indicated on the right side of the figure. The radioactive tantalum isotopes were produced using fusion evaporation reactions (FE), light-particle reactions (LP), neutron capture reactions (NC), photo-nuclear reactions (PN), spallation (SP), and projectile fragmentation or fission (PF). The stable isotope was identified using mass spectroscopy (MS). Light particles also include neutrons produced by accelerators. The discovery of each tantalum isotope is discussed in detail and a summary is presented in Table 1.

\subsection*{$^{155}$Ta}\vspace{0.0cm}

Page et al. announced the discovery of $^{155}$Ta in the 2007 paper ``$\alpha$ decay of $^{159}$Re and proton emission from $^{155}$Ta'' \cite{2007Pag01}. Isotopically enriched $^{106}$Cd was bombarded with a 300 MeV $^{58}$Ni beam at the Jyv\"askyl\"a accelerator laboratory forming $^{159}$Re in the (p4n) fusion-evaporation reaction. $^{155}$Ta was observed following the $\alpha$-decay of $^{159}$Re with the RITU gas-filled separator and the GREAT spectrometer. ``This $\alpha$ decay populates a state in the closed neutron shell nucleus $^{155}$Ta, which decays by emitting 1444$\pm$15 keV protons with a half-life of 2.9$^{+1.5}_{-1.1}$ ms. These values are consistent with the emission of the proton for a $\pi\emph{h}_{11/2}$ orbital. These results fit in with the systematics of proton and $\alpha$-particle separation energies in the region, but disagree with the previously reported decay properties of $^{155}$Ta.'' This half-life corresponds to the currently accepted value. The disagreement mentioned in the quote refers to a previous measurement of E$_p$ = 1765(10)~keV and T$_{1/2}$ = 12$^{+4}_{-3}\mu$s \cite{1999Uus01} which was evidently incorrect.

\subsection*{$^{156}$Ta}\vspace{0.0cm}

Page et al. discovered $^{156}$Ta as reported in the 1992 paper ``Discovery of new proton emitters $^{160}$Re and $^{156}$Ta'' \cite{1992Pag01}. A 300~MeV $^{58}$Ni beam from the Daresbury tandem accelerator bombarded an enriched $^{106}$Cd target forming $^{160}$Re in the fusion-evaporation reaction $^{106}$Cd($^{58}$Ni,p3n) and $^{156}$Ta was populated by subsequent $\alpha$-decay. Residues were separated using the Daresbury Recoil Mass Separator and charged particles were measured with a double-sided silicon strip detector. ``The half-life of this correlated decay line was determined as 165$^{+165}_{-55}$ ms which is significantly shorter than the value of ~1 s predicted for beta decay, the principal competing decay mode for $^{156}$Ta. This new decay line is therefore assigned to the proton decay of $^{156}$Ta and a $\emph{Q}$ value of 1028$\pm$13 keV was deduced for these correlated proton decays.'' This half-life agrees with the currently accepted value of 144(24)~ms. In 1989 Hofman et al. reported a lower limit of 10~ms for the $\beta$-decay half-life of a high-spin state of $^{156}$Ta \cite{1989Hof01}.

\subsection*{$^{157-161}$Ta}\vspace{0.0cm}

``Alpha decay studies of very neutron deficient isotopes of Hf, Ta, W, and Re'' was published in 1979 by Hofmann et al. describing the observation of $^{157}$Ta, $^{158}$Ta, $^{159}$Ta, $^{160}$Ta, and $^{161}$Ta, \cite{1979Hof01}. Targets of $^{103}$Rh, $^{nat,108,110}$Pd, and $^{107,109}$Ag were bombarded with beams of $^{58}$Ni from the GSI UNILAC linear accelerator. Evaporation residues were separated with the high-velocity SHIP separator. ``In the investigated reactions the eleven new isotopes $^{161-164}$Re, $^{160}$W, $^{157-161}$Ta, and $^{156}$Hf could be identified.'' The reported half-lives of 5.3(18)~ms ($^{157}$Ta), 36.8(16)~ms ($^{158}$Ta), and 570(180)~ms ($^{159}$Ta) are consistent with the presently adopted values 4.3(1)~ms, 36.0(8)~ms, and 514(9)~ms for isomeric states, respectively. For $^{160}$Ta and $^{161}$Ta only the $\alpha$-decay energies were measured.

\subsection*{$^{162,163}$Ta}\vspace{0.0cm}

In ``On-line separation of $^{163}$Ta and $^{162}$Ta'' Liang et al. discovered $^{162}$Ta and $^{163}$Ta in 1985 \cite{1985Lia01}. A $^{175}$LuF$_{3}$ powder target was bombarded by a 280~MeV $^3$He beam. $^{163}$Ta and $^{162}$Ta were separated with the Isocele-2 on-line separator and X- and $\gamma$-ray spectra were measured. ``By on-line selective separation on molecular fluoride compounds at the ISOCELE 2 facility, two new isotopes: $^{163}$Ta (T$_{1/2}$=10.5$\pm$1.8 s) and $^{162}$Ta (T$_{1/2}$=3.5$\pm$0.2 s) have been identified.'' These half-lives agree with the currently accepted values of 3.57(12)~s and 10.6(18)~s for $^{162}$Ta and $^{163}$Ta, respectively.

\subsection*{$^{164}$Ta}\vspace{0.0cm}

In the 1982 paper ``Untersuchung des Neuen Isotops $^{164}$Ta'', Eichler et al. reported the first discovery of $^{164}$Ta \cite{1982Eic01}. An enriched $^{151}$Eu target was bombarded with 154 and 164~MeV $^{20}$Ne beams from the Dubna U-300 heavy-ion accelerator and $^{164}$Ta was produced in the fusion-evaporation reaction $^{151}$Eu($^{20}$Ne,7n). Decay curves and $\gamma$-ray spectra were measured following chemical separation. ``In periodic batch--wise experiments the Ta-fraction was chemically separated and the isotope $^{164}$Ta with a half-life of 20$\pm$5 s was found, emitting 210.6 keV $\gamma$-rays. The assignment is based on the measured growth and decay curve of $^{164}$Hf.'' This half-life is consistent with the presently accepted value of 14.2(3)~s. Less than four months later Liang et al. independently reported a half-life of 13.7(6)~s \cite{1982Lia01}.

\subsection*{$^{165}$Ta}\vspace{0.0cm}

``Untersuchung der Produkte der Reaktion $^{151}$Eu + $^{20}$Ne'' was published in 1982, announcing the discovery of $^{165}$Ta by Bruchertseifer et al. \cite{1982Bru01}. An enriched $^{151}$Eu target was bombarded with 154 and 164~MeV $^{20}$Ne beams from the Dubna U-300 heavy-ion accelerator and $^{165}$Ta was produced in the fusion-evaporation reaction $^{151}$Eu($^{20}$Ne,6n). Decay curves and $\gamma$-ray spectra were measured following chemical separation. ``In the Ta fraction $^{165}$Ta was identified. Its half-life is 35$\pm$10 s.'' This half-life value is in agreement with the presently accepted value of 31.0(15)~s. Less than nine months later Liang et al. independently reported a half-life of 31.0(15)~s \cite{1982Lia01}.

\subsection*{$^{166}$Ta}\vspace{0.0cm}

In the paper ``Identification of a new isotope: $^{166}$Ta,'' Leber et al. reported the discovery of $^{166}$Ta in 1977 \cite{1977Leb01}. A 147~MeV $^{16}$O beams from the Yale heavy-ion accelerator bombarded self-supporting terbium targets and $^{166}$Ta was produced in the fusion-evaporation reaction $^{159}$Tb($^{16}$O,9n). Recoil nuclei were thermalized in a helium chamber, then transported with a helium jet to a magnetic tape where $\gamma$ rays were measured with a Ge(Li) detector. ``Excitation function measurements for the production of $^{170}$Ta and $^{166}$Ta via the $^{159}$Ta($^{16}$0,Xn) reaction indicated that the $^{166}$Ta yield would be maximized at $\sim$145$-$150~MeV. At this bombarding energy, $\gamma$-ray transitions at 158.7 and 311.7~keV were observed to peak in intensity. These transitions had been previously reported to result from dexcitation of the 4$^+$ and 2$^+$ levels of the ground state band in $^{166}$Hf. Observation of these transitions at the bombarding energy which was expected to maximize the yield of $^{166}$Ta indicated that the transitions resulted from the production of $^{166}$Ta and its subsequent decay to $^{166}$Hf.''  The measured half-life of 32(3)~s agrees with the currently adopted value of 34.4(5)~s.

\subsection*{$^{167}$Ta}\vspace{0.0cm}

The 1982 paper ``Selective on-line separation of new Ta, Zr and Sr isotopes'' by Liang et al. described the discovery of $^{167}$Ta \cite{1982Lia01}. A 280~MeV  $^3$He beam from the IPN Orsay synchrocyclotron bombarded lutetium metal and LuF$_3$ powder. $^{167}$Ta was identified with the Isocele-2 on-line separator. ``The measured half-life of 1.4 $\pm$ 0.3 mn differs from the value 2.9 $\pm$ 0.15 mn previously reported for this isotope.'' This value agrees with the currently accepted value of 80(4)~s. The previously reported value of 2.9(15)~min \cite{1969Arl03} was evidently incorrect.

\subsection*{$^{168-171}$Ta}\vspace{0.0cm}

In 1969, the discovery of $^{168}$Ta, $^{169}$Ta, $^{170}$Ta, and $^{171}$Ta was announced in ``New neutron-deficient isotopes of tantalum with mass numbers from 167 to 171, and the systematics of the half lives of deformed neutron-deficient nuclei with 150 $<$ A $<$ 190'' by Arlt et al. \cite{1969Arl01}. HgO and HReO$_{4}$ targets were bombarded with 660~MeV protons from the Dubna synchrocyclotron. Gamma-ray spectra were measured with a Ge(Li) detector following chemical separation. ``The $^{171}$Ta was identified from its genetic relations to the daughter hafnium and lutetium isotopes... [The figure] shows decay curves for $^{171}$Ta constructed from the intensities of the 662 keV ($^{171}$Hf) and 741 ($^{171}$Lu) $\gamma$ ray from Hf preparations separated from the Ta fraction at 20 min intervals. The half life of $^{171}$Ta is 25$\pm$2 min... To identify lighter tantalum isotopes we undertook four series of experiments with hafnium preparations separated at 4 to 6 min. intervals from the tantalum fraction of targets bombarded from 3 to 15 min. The results gave a half life of 7.0$\pm$0.5~min for $^{170}$Ta... The presence of $^{169}$Ta was shown in the same way from the daughter isobars $^{169}$Lu and $^{169}$Yb; its half life is 5.0$\pm$0.5~min... We observed the 239 keV $^{167}$Lu $\gamma$ rays and the 106, 113 and 176 keV $^{167}$Yb ones, and from their intensities from hafnium preparations separated at 4 min intervals we obtained the half life of $^{167}$Ta as 2.9$\pm$1.5 min.''
These half-lives agree with the presently accepted values of 2.0(1)~min, 4.9(4)~min, 6.76(6)~min, and 23.3(3)~min, for $^{168}$Ta, $^{169}$Ta, $^{170}$Ta, and $^{171}$Ta, respectively. A year later Rezanka et al. reported the observation of the new isotopes $^{170}$Ta and $^{171}$Ta \cite{1970Rez01} apparently unaware of the work by Arlt et al.\ \cite{1969Arl01}.

\subsection*{$^{172}$Ta}\vspace{0.0cm}

In the 1964 paper ``P\'eriode du premier \'etat excit\'e du noyau de hafnium 172'' Abou-Leila reported the observation of $^{172}$Ta \cite{1964Abo01}. Hafnium oxide was bombarded with 72 MeV protons from the Orsay synchrocyclotron. $^{172}$Ta was isotopically separated with a two stage separator and gamma rays were measured with a NaI(Tl) detector. ``Nous avons trouv\'e la p\'eriode de $^{172}$Ta \'egale \`a 44$\pm$1 mn en d\'esaccord avec la valeur obtenue par Butement (23.6$\pm$1.2mn).'' [We found a half-life of 44$\pm$1~min for $^{172}$Ta in disagreement with the value obtained by Butement (23.6$\pm$1.2~min)]. The measured half-life of 44(1)~min agrees with the currently adopted value of 36.8(3)~min. We credit Abou-Leila with the discovery because he identified the first excited state of the daughter nucleus ($^{172}$Hf) correctly and the somewhat longer half-life was later explained by a possible contamination \cite{1972Cha01}. The result by Butement and Briscoe \cite{1961But01} mentioned in the quote was not considered correct by neither Abou-Leila nor later by Chang and Cheney \cite{1964Abo01,1972Cha01}.

\subsection*{$^{173-175}$Ta}\vspace{0.0cm}

Faler and Rasmussen described the observation of $^{173}$Ta, $^{174}$Ta, and $^{175}$Ta in ``New neutron-deficient isotopes of tantalum'' in 1960 \cite{1960Fal01}. $^{14}$N beams between 35 and 95~MeV from the Berkeley heavy-ion linear accelerator bombarded Ho$_2$O$_3$ powder targets and (xn) fusion evaporation reactions produced tungsten which subsequently decayed to tantalum isotopes. Decay curves and $\gamma$-ray spectra were measured with an end-window G-M counter and a Na(Tl) detector following chemical separation. ``Bombardment of Ho$_{2}$O$_{4}$ with N$^{14}$ ions in the Berkeley heavy-ion linear accelerator has resulted in the discovery of new isotopes of tantalum which have been assigned as Ta$^{173}$ and Ta$^{174}$. They have half-lives of 3.7~hr and 1.3~hr, respectively. Tantalum-172 was not observed and is believed to have a half-life shorter than 30 minutes. Gamma-ray spectra have been obtained for these two isotopes and for Ta$^{175}$. Tantalum-175, with an 11-hr half-life, has also been produced by 48-Mev alpha-particle bombardment of Lu$_2$O$_3$, and its conversion-electron spectrum was studied.'' These half-lives agree with the presently adopted values of 3.14(13)~h, 1.14(8)~h, and 10.5(2)~h, for $^{173}$Ta, $^{174}$Ta, and $^{175}$Ta, respectively. $^{175}$Ta was not considered a new observation quoting a presentation at a meeting of the American Chemical Society. An 11~h half-life was also mentioned in a note added in proof in a paper by Harmatz et al. \cite{1959Har01}, which in turn was quoted by Grigorev who observed several $\gamma$-rays with a half-life between 8 and 11~h. They concluded that $^{175}$Ta as well as $^{176}$Ta were present in their data \cite{1960Gri01}.

\subsection*{$^{176,177}$Ta}\vspace{0.0cm}

Wilkinson and Hicks reported the first observation of $^{176}$Ta and $^{177}$Ta in the 1948 paper ``Some new radioactive isotopes of Tb, Ho, Tm Lu, Ta, W and Re'' \cite{1948Wil02}. The Berkeley 60-in cyclotron was used to bombard lutetium with 20 and 38~MeV $\alpha$-particles and hafnium an tantalum with 19~MeV deuterons. Absorption measurements were performed and decay curves recorded following chemical separation. The results were only summarized in a table. The measured half-lives of 8.0~h ($^{176}$Ta) and 2.66~d ($^{177}$Ta) agree with the currently accepted values of 8.09(5)~h and 56.56(6)~h, respectively.

\subsection*{$^{178,179}$Ta}\vspace{0.0cm}

Wilkinson discovered $^{178}$Ta and $^{179}$Ta as reported in the paper ``Neutron deficient radioactive isotopes of tantalum and wolfram'' in 1950 \cite{1950Wil01}. The Berkeley 60-in cyclotron was used to bombard lutetium with 20, 30, and 38~MeV $\alpha$-particles and hafnium with 10 MeV protons. Decay curves were measured following chemical separation. ``9.35$\pm$0.03-min. Ta$^{178}$: ... The 9.3-min. activity has been observed directly in the bombardment of hafnium with 10 Mev protons. Allocation of the 21.5-day wolfram parent is made to mass 178 on the basis of reaction yields; since a longerlived tantalum has been allocated to mass 178 from Lu+$\alpha$ bombardments the 9.35 min. activity must be an isomer decaying independently... $\sim$600~day Ta$^{179}$: ... The isotope is not formed by decay of wolfram parents of half-life greater than one hour, and since no short-lived tantalum daughters of the 30-min. wolfram activity definitely allocated to mass 179 have been found, it is most likely that the 600-day activity has mass 179.'' These half-lives agree with the presently adopted values of 9.31(3)~min and 1.82(3)~y for $^{178}$Ta and $^{179}$Ta, respectively.

\subsection*{$^{180}$Ta}\vspace{0.0cm}

``Artificial radioactivity of tantalum'' was published in 1938 by Oldenberg, describing the observation of $^{180}$Ta \cite{1938Old01}. Tantalum was bombarded with fast neutrons produced by irradiating lithium with 5.5~MeV deuterons from the Berkeley cyclotron. Decay curves were measured with a Lauritsen electroscope following chemical separation. ``Fast neutron bombardment excites, in addition, an 8.2~hour period with the emission of electrons, K radiation of Ta, and $\gamma$-rays. The process responsible for these effects is probably the capture of one neutron with the ejection of two neutrons. The product nucleus, Ta$^{180}$, goes over to Hf$^{180}$ largely by K electron capture; in this process either $\gamma$~rays are emitted or by their internal conversion extranuclear electrons ejected.'' This half-life agrees with the presently adopted value of 8.152(6)~h. Previously, Pool et al. reported a half-life without a mass assignment \cite{1937Poo01} and the 20~min half-life measured by Bothe and Gentner \cite{1937Bot03} was evidently incorrect.

\subsection*{$^{181}$Ta}\vspace{0.0cm}

In 1932 Aston discovered the only stable isotope of tantalum, $^{181}$Ta, as reported in ``Constitution of tantalum and niobium'' \cite{1932Ast01}. Tantalum penta-fluoride was volatilised in the discharge tube of the Cavendish mass spectrometer. ``Tantalum, which was investigated first, gave a strong line at 181 followed by a diminishing series 200, 219... due to TaF, TaF$_{2}$... Neither the expected isotope 183 nor any other could be detected even to one-fiftieth of the main line.''

\subsection*{$^{182}$Ta}\vspace{0.0cm}

``Artificial radioactivity of tantalum'' was published in 1938 by Oldenberg, describing the observation of $^{182}$Ta \cite{1938Old01}. Tantalum was bombarded with slow neutrons at the Berkeley Radiation Laboratory. Decay curves were measured with a Lauritsen electroscope. ``The long period of 200 days $\pm$100 given by Fomin and Houtermans was confirmed. A more accurate value of the half-life is 97$\pm$8 days. As there exists only one stable isotope, Ta$^{181}$ capture of slow neutrons must lead to Ta$^{182}$.'' This half-life is close to the presently adopted value of 114.74(12)~d. In the reference by Fomin and Houtermans mentioned in the quote no mass assignment was made \cite{1936Fom01}.

\subsection*{$^{183}$Ta}\vspace{0.0cm}

$^{183}$Ta was first observed by Butement in 1950 as reported in ``New radioactive isotopes produced by nuclear photo-disintegration'' \cite{1950But01}. $^{183}$Ta was produced through irradiation of tungstic acid by 23~MeV x-rays from the synchrotron in the photonuclear reaction $^{184}$W($\gamma$,p) and chemically separated from other resultant isotopes \cite{1951But01}. In the original paper \cite{1950But01} a probable assignment was only given in a table. More details were reported in the subsequent publication \cite{1951But01}: ``The yields of 48-minute and 116-hour tantalums were in the ratio of 1:1.2 respectively, which is in conformity with the nearly equal abundances of $^{184}$W and $^{185}$W. Probably the 48-minute activity is $^{185}$Ta and the 116-hour activity $^{183}$Ta.'' The half-life measured for $^{183}$Ta agrees with the presently adopted value of 5.1(1)~d.

\subsection*{$^{184}$Ta}\vspace{0.0cm}

Butement and Poe announced the discovery of $^{184}$Ta in their 1955 paper ``Radioactive $^{184}$Tantalum'' \cite{1955But02}. Fast neutrons produced by bombarding beryllium with 20 MeV protons were used to irradiate tungstic acid targets. Decay curves and $\beta$- and $\gamma$-spectra were measured following chemical separation. ``The best values for the half-life of 8.7$\pm$0.1~h were obtained from five sources the decay of which was followed by counting those beta-particles which passed through 228~mg/cm$^2$ of aluminium, this being sufficient to absorb all the beta-particles from longer-lived activity present. The long-lived background activity was then reduced to that due to inefficiently counted gamma-rays. The mass assignment of the 8.7 h activity was made by the use of tungstic acid enriched in the tungsten isotope of mass 184.'' This half-life corresponds to the presently accepted value.

\subsection*{$^{185}$Ta}\vspace{0.0cm}

$^{185}$Ta was first observed by Butement in 1950 as reported in ``New radioactive isotopes produced by nuclear photo-disintegration'' \cite{1950But01}. $^{185}$Ta was produced through irradiation of tungstic acid by 23~MeV x-rays from the synchrotron in the photonuclear reaction $^{186}$W($\gamma$,p) and chemically separated from other resultant isotopes \cite{1951But01}. In the original paper \cite{1950But01} the probably assignment was only given in a table. More details were reported in the subsequent publication \cite{1951But01}: ``The yields of 48-minute and 116-hour tantalums were in the ratio of 1:1.2 respectively, which is in conformity with the nearly equal abundances of $^{184}$W and $^{185}$W. Probably the 48-minute activity is $^{185}$Ta and the 116-hour activity $^{183}$Ta.'' The half-life measured for $^{185}$Ta agrees with the presently adopted value of 49.4(15)~min.

\subsection*{$^{186}$Ta}\vspace{0.0cm}

``Radioactive $^{186}$Tantalum'' was published by Po\"e in 1955, describing the first observation of $^{186}$Ta \cite{1955Poe01}. Tungstic acid targets were irradiated with fast neutrons produced by bombarding beryllium with protons from the Harwell cyclotron. Decay curves and $\beta$- and $\gamma$-ray spectra were measured following chemical separation. ``The new radioactive isotope $^{186}$Ta has been prepared and its decay characteristics determined as follows: Half-life: 10.5$\pm$0.5~min, Maximum $\beta$-energy: 2.2$_2$~MeV, Conversion-electron energies: $\le$0.15~MeV, $\gamma$-ray energies: 125, 200, 300, 410, 510, 610, 730, 940 and possibly $\sim$1150~keV. The mass assignment suggested by its production by (n,p), but not by ($\gamma$,p), reactions on tungsten was confirmed by experiments using tungstic acid enriched in $^{186}$W.'' This half-life agrees with the presently accepted value of 10.5(3)~min.

\subsection*{$^{187-189}$Ta}\vspace{0.0cm}

Benlliure et al. published the discovery of $^{187}$Ta, $^{188}$Ta, and $^{189}$Ta in the 1999 paper entitled ``Production of neutron-rich isotopes by cold fragmentation in the reaction $^{197}$Au + Be at 950 \emph{A} MeV'' \cite{1999Ben01}. A 950~A$\cdot$MeV $^{197}$Au beam from the SIS synchrotron of GSI was incident on a beryllium target and $^{187}$Ta, $^{188}$Ta, and $^{189}$Ta were produced in projectile fragmentation reactions. The FRS fragment separator was used to select isotopes with a specific mass-to-charge ratio.  ``In the right part of [the figure] the projected A/Z distributions are shown for the different elements transmitted in this setting of the FRS. In this setting the isotopes $^{193}$Re, $^{194}$Re, $^{191}$W, $^{192}$W, $^{189}$Ta, $^{187}$Hf and $^{188}$Hf were clearly identified for the first time. Only isotopes with a yield higher than 15 counts were considered as unambiguously identified.'' The abstracts incorrectly lists $^{189}$Tl instead of $^{189}$Ta as the discovery of a new isotope. Although not explicitly mentioned in the text clear evidence for the presence of $^{187}$Ta and $^{188}$Ta can be seen in the A/Z isotopic identification plot. It is not clear why Benlliure et al. did not consider $^{187}$Ta and $^{188}$Ta as new isotopes, because no previous publications reporting the observation of these nuclei could be found.

\subsection*{$^{190-192}$Ta}\vspace{0.0cm}

Alkhomashi et al. observed $^{190}$Ta, $^{191}$Ta, and $^{192}$Ta in the 2009 paper ``$\beta^-$-delayed spectroscopy of neutron-rich tantalum nuclei: Shape evolution in neutron-rich tungsten isotopes'' \cite{2009Alk01}. A beryllium target was bombarded with a 1 GeV/nucleon $^{208}$Pb beam from the SIS-18 heavy-ion synchrotron at GSI, Germany. Projectile-like fragments were separated with the FRS and implanted in a series of double-sided silicon strip detectors where correlated $\beta$-decay was measured in coincidence with $\gamma$-rays in the $\gamma$-ray spectrometer RISING. ``The insets of [the figure] show the time spectra associated with $\beta$ decays of $^{188}$Ta, $^{190}$Ta, and $^{192}$Ta, gated on discrete $\gamma$-ray lines identified in the tungsten daughter nuclei.'' The reported half-lives of 5.3(7)~s and 2.2(7)~s correspond to the presently adopted values of $^{190}$Ta and $^{192}$Ta, respectively. Although not specifically mentioned in the text, evidence for $^{191}$Ta is clearly visible in the two-dimensional particle identification plot. The authors did not consider their observation a new discovery because of a previous publication in a conference proceeding \cite{2009Ste01}.

\section{Discovery of $^{159-196}$Re}

Thirty-eight rhenium isotopes from A = 159--196 have been discovered so far; these include 2 stable, 26 neutron-deficient and 10 neutron-rich isotopes. According to the HFB-14 model \cite{2007Gor01} $^{251}$Re should be the last particle stable neutron-rich nucleus. The proton dripline has already been crossed with the observation of the proton emitters $^{159}$Re and $^{160}$Re; however, about six additional proton-rich rhenium isotopes could still have half-lives longer than 10$^{-9}$~s \cite{2004Tho01}. Thus, about 61 isotopes have yet to be discovered corresponding to 62\% of all possible rhenium isotopes.

\begin{figure}
	\centering
	\includegraphics[scale=.7]{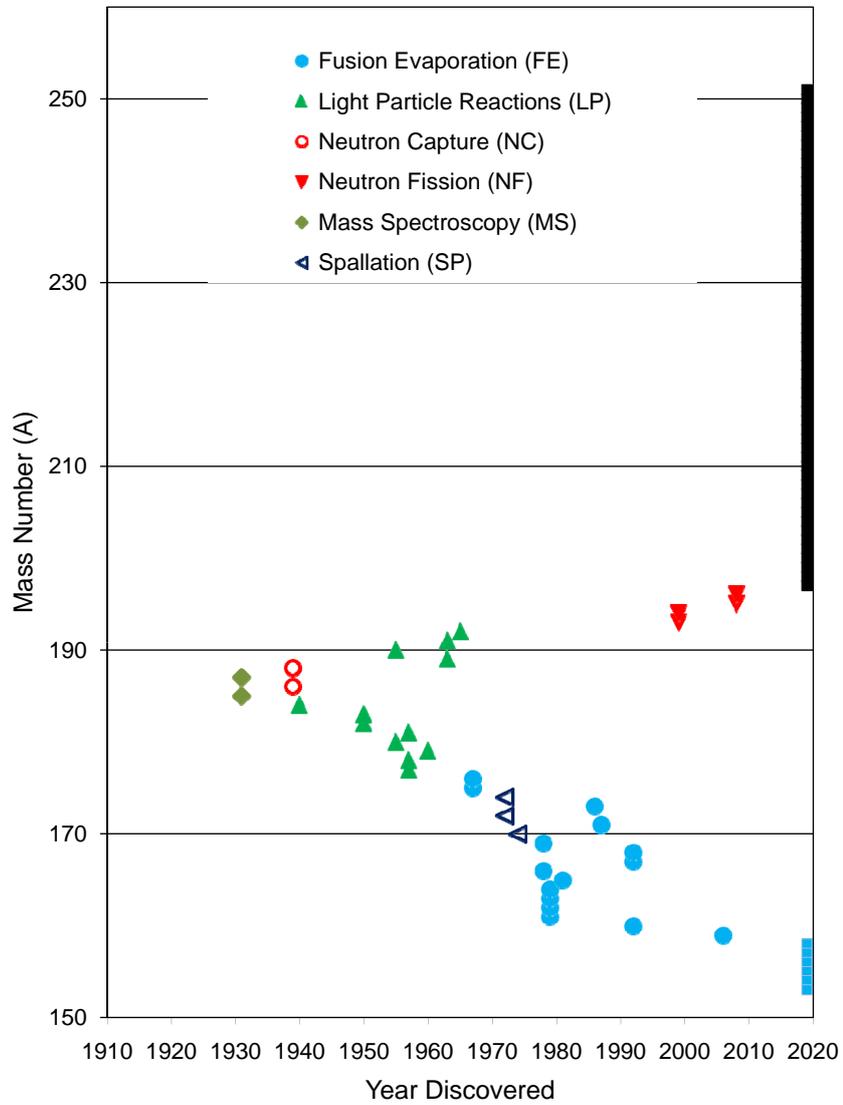}
	\caption{Rhenium isotopes as a function of time when they were discovered. The different production methods are indicated. The solid black squares on the right hand side of the plot are isotopes predicted to be bound by the HFB-14 model. On the proton-rich side the light blue squares correspond to unbound isotopes predicted to have half-lives larger than $\sim 10^{-9}$~s.}
\label{f:year-re}
\end{figure}

Figure \ref{f:year-re} summarizes the year of first discovery for all rhenium isotopes identified by the method of discovery. The range of isotopes predicted to exist is indicated on the right side of the figure. The radioactive rhenium isotopes were produced using fusion evaporation reactions (FE), light-particle reactions (LP), neutron capture reactions (NC), spallation reactions (SP), and projectile fission or fragmentation (PF). The stable isotopes were identified using mass spectroscopy (MS). Light particles also include neutrons produced by accelerators. In the following, the discovery of each rhenium isotope is discussed in detail.

\subsection*{$^{159}$Re}\vspace{0.0cm}

``Probing the limit of nuclear existence: Proton emission from $^{159}$Re'' reported the discovery of $^{159}$Re in 2006 by Joss et al. \cite{2006Jos01}. An enriched $^{106}$Cd target was bombarded with a 300~MeV $^{58}$Ni beam at the Jyv\"askyl\"a Accelerator Laboratory and $^{159}$Re was produced in the fusion-evaporation reaction $^{106}$Cd($^{58}$Ni,p4n). Reaction products were identified using the RITU gas-filled separator and the GREAT focal-plane spectrometer. ``The 1.8~MeV peak is assigned as the proton decay from the previously unknown nuclide $^{159}$Re. This yield corresponds to one $^{159}$Re nucleus in every 4 million evaporation residues implanted into GREAT. The few counts at higher energy represent real correlations with the 6600$\pm$3 keV $\alpha$ decay of $^{162}$Os populated directly as an evaporation residue. The half-life of the $^{159}$Re proton decay peak was measured as 21$\pm$4~$\mu$s using the maximum likelihood method.'' This half-life was used in the calculation of the currently adopted value of 20(4)~$\mu$s.

\subsection*{$^{160}$Re}\vspace{0.0cm}

Page et al. discovered $^{160}$Re as reported in the 1992 paper ``Discovery of new proton emitters $^{160}$Re and $^{156}$Ta'' \cite{1992Pag01}. A 300~MeV $^{58}$Ni beam from the Daresbury tandem accelerator bombarded an enriched $^{106}$Cd target and $^{160}$Re was formed in the fusion-evaporation reaction $^{106}$Cd($^{58}$Ni,p3n). Residues were separated using the Daresbury Recoil Mass Separator and charged particles were measured with a double-sided silicon strip detector. ``Combining the data from both the proton and the alpha decay branches, a value of 790$\pm$160~$\mu$s was obtained for the half-life of $^{160}$Re using the method described in [the reference].'' This half-life agrees with the currently accepted value of 0.82$^{+15}_{-9}$~ms.

\subsection*{$^{161-164}$Re}\vspace{0.0cm}

``Alpha decay studies of very neutron deficient isotopes of Hf, Ta, W, and Re'' was published in 1979 by Hofmann et al. describing the observation of $^{161}$Re, $^{162}$Re, $^{163}$Re, and $^{164}$Re \cite{1979Hof01}. Targets of $^{103}$Rh, $^{nat,108,110}$Pd, and $^{107,109}$Ag were bombarded with beams of $^{58}$Ni from the GSI UNILAC linear accelerator. Evaporation residues were separated with the high-velocity SHIP separator. ``In the investigated reactions the eleven new isotopes $^{161-164}$Re, $^{160}$W, $^{157-161}$Ta, and $^{156}$Hf could be identified.'' The reported half-lives of 10$^{+15}_{-5}$~ms ($^{161}$Re), 100(30)~ms ($^{162}$Re), 260(40)~ms ($^{163}$Re), and 0.9(7)~s ($^{164}$Re) are consistent with the presently adopted values 15.6(9)~ms, 107(13)~ms, 214(5)~ms (isomeric state), and 520(230)~ms, respectively.

\subsection*{$^{165}$Re}\vspace{0.0cm}

The 1981 paper ``New neutron deficient isotopes in the range of elements Tm to Pt'' reported the discovery of $^{165}$Re by Hofmann et al. at the GSI linear accelerator UNILAC \cite{1981Hof01}. $^{165}$Re was produced in reactions bombarding neutron deficient targets between Mo and Sn with a beam of $^{58}$Ni as well as bombarding targets between V and Ni with a beam of $^{107}$Ag, with energies between 4.4 MeV/u and 5.9 MeV/u. Residues were separated using the velocity filter SHIP. ``An $\alpha$ line of 5506~keV was observed as daughter in the $^{169}$Ir decay and is assigned to $^{165}$Re.'' The measured half-life of 2.4(6)~s agrees with the currently accepted value of 2.61$^{+14}_{-13}$~s.

\subsection*{$^{166}$Re}\vspace{0.0cm}

Schrewe et al. reported the discovery of $^{166}$Re in the 1978 paper ``Evidence of new isotopes: $^{169,170}$Ir, $^{166,167,168}$Re'' \cite{1978Sch01}. $^{93}$Nb and $^{89}$Y targets were bombarded with 5.6~MeV/u and 6.8~MeV/u $^{84}$Kr beams from the GSI UNILAC linear accelerator. Recoil products were collected on a rotating wheel which transported the activities in front of three Si-surface barrier detectors measuring $\alpha$ particles. ``Using three alpha detectors and spectrum multiscaling, energies and half-lives were measured. Ir and Re isotopes were identified by cross bombardments, excitation function data and $\alpha$-systematics. The decay characteristics of the new species are as follows:... $^{166}$Re E$_\alpha$ = 5.495$\pm$0.010~MeV, T$_{1/2}$ = 2.2$\pm$0.4~s.'' This half-life is consistent with the currently accepted value of 2.25(21)~s.

\subsection*{$^{167,168}$Re}\vspace{0.0cm}

The identification of $^{167}$Re and $^{168}$Re was published in 1992 in ``Revision of the decay data of $^{166-170}$Re, including new isomers $^{167m,169m}$Re'' by Meissner et al. \cite{1992Mei01}. $^{141}$Pr targets were irradiated with a 235~MeV $^{32}$Si beam from the VICKSI accelerator facility at the Hahn-Meitner-Institut, Berlin, Germany. The reaction products were transported to a surface barrier $\alpha$-detector, mounted between a $\gamma$-X detector and a $\gamma$-detector with a helium jet system and a fast transport tape. ``The earlier reported 5.26~MeV and also the new 5.02 MeV $\alpha$-rays display a similar excitation function as $^{167}$W and are consequently assigned to $^{167}$Re... From the excitation functions in [the figure], the new 4.83~MeV $\alpha$-radiation has to be assigned to the isotope $^{168}$Re.'' The reported half-life of 5.7(14)~s for $^{167}$Re agrees with the presently adopted value of 5.9(3)~s and the half-life of 4.4(1)~s for $^{168}$Re corresponds to the current value. The 5.26~MeV $\alpha$-energy mentioned in the quote for $^{167}$Re had previously been assigned to $^{168}$Re \cite{1978Cab01,1982Del02,1984Sch01}. Meissner also demonstrated that $\alpha$-decay assignments by Schrewe et al. \cite{1984Sch01} to $^{166-168}$Re were most likely from $^{163-165}$W.

\subsection*{$^{169}$Re}\vspace{0.0cm}

``Copper ion induced reactions on $^{110-108-106}$Cd, $^{109-107}$Ag and $^{110}$Pd. New rhenium, osmium and iridium isotopes'' was published in 1978 by Cabot et al. announcing the discovery of $^{169}$Re \cite{1978Cab01}. A 400~MeV $^{63}$Cu beam from the ALICE accelerator at Orsay, France, bombarded isotopically enriched $^{108}$Cd, $^{109}$Ag, and $^{110}$Pd targets to populate $^{169}$Re in the fusion-evaporation reactions ($^{63}$Cu,2p), ($^{63}$Cu,p2n), and ($^{63}$Cu,4n), respectively. Alpha particles from fragments collected by a He-jet were detected to determine the decay energies and half-lives. ``Our conclusion is that the 5.05~MeV emission is due to the $^{169}$Re $\alpha$-decay. Then $^{169}$Re is the first identified $\alpha$-emitter of this element.''

\subsection*{$^{170}$Re}\vspace{0.0cm}

In 1974, Berlovich et al. reported the discovery of $^{170}$Re in the paper ``Decay of the new isotopes $^{170}$Re and $^{172}$Re'' \cite{1974Ber01}. A thallium sulfate solution was irradiated with a 1-GeV proton beam. Gamma-ray spectra were measured following chemical separation. ``[The table] lists the gamma lines for which the half-life is 9$\pm$2~sec. The 156.6 keV line has the same energy as the $2^+ \rightarrow 0^+$ transition of the main rotational band of $^{170}$W, which is the basis for associating this half-life with $^{170}$Re.'' This half-life agrees with the presently adopted value of 9.2(2)~s. Later in the year Sterna et al. reported independently a half-life of 8.0(5)~s \cite{1975Ste01}.

\subsection*{$^{171}$Re}\vspace{0.0cm}

In 1987, Runte et al. identified $^{171}$Re in their paper ``The decay of the new isotope $^{171}$Re'' \cite{1987Run02}. A 239~MeV $^{36}$Ar beam from the Hahn-Meitner-Institut VICKSI accelerator facility bombarded $^{139}$LaF$_3$ targets. Reaction products were transported to a helium chamber containing NaCl aerosols and then sprayed onto a transport tape between two germanium detectors, where $\gamma$-ray spectra were measured. ``The comparison of the excitation functions leads to the conclusion that the emitter of the new activity is generated in a four particle evaporation reaction. The observation of coincidences with W-X-rays and annihilation radiation pins down the element rhenium and therefore proves the identification of the new isotope $^{171}$Re.'' The reported half-life of 15.2(4)~s corresponds to the currently accepted value. About six months earlier, Szymanski et al. reported an upper limit of 20~s for the half-life of $^{171}$Re \cite{1987Szy01}.

\subsection*{$^{172}$Re}\vspace{0.0cm}

In the 1972 paper ``Short-lived osmium isotopes,'' Berlovich et al. reported the observation of $^{172}$Re \cite{1972Ber01}. A mercury target was bombarded with 1 GeV protons from the Leningrad synchrocyclotron. Gamma-ray spectra were measured with a Ge(Li) detector following chemical separation. ``We also observe a transition with E$_\gamma$ = 254 keV and T$_{1/2}$ = (0.8$\pm$0.2) min which presumably refers to the decay of $^{172}$Re.'' This half-life agrees with the presently accepted value for the isomeric state of 55(5)~s.

\subsection*{$^{173}$Re}\vspace{0.0cm}

In 1986, Szymanski et al. reported the existence of $^{173}$Re in their paper entitled ``Absolute gamma ray abundances of rhenium and tungsten isotopes: Part III, A=173'' \cite{1986Szy01}. Thin holmium foils were bombarded with a 151.2~MeV $^{16}$O beam from the Manchester heavy-ion linear accelerator and $^{173}$Re was formed in the fusion-evaporation reaction $^{169}$Ho($^{16}$O,8n). Reaction products were removed from the target area with a helium jet recoil transportation system and $\gamma$-ray spectra were measured. ``Three possible candidates for $^{173}$Re photons emerged here, at 181.5, 190.7 and 373.6 keV. They were assigned on the basis of excitation function and half-life measurements... All three $\gamma$-rays were observed in two experiments, R1 and R2. The decay parameters are given in [the table]. Due to the longer irradiation counting scheme of R2, the peaks seen were of lower statistical accuracy and had fewer data points. The half-lives measured were consistent for both sets of values and the value proposed here is the mean of the six, i.e.\ t$_{1/2}$($^{173}$Re) = 1.98$\pm$0.26 minutes.'' This half-life corresponds to the currently adopted value. Previously a half-life of 0.6~min was assigned to either $^{172}$Re or $^{173}$Re \cite{1973Ber01}.

\subsection*{$^{174}$Re}\vspace{0.0cm}

In the 1972 paper ``Short-lived osmium isotopes,'' Berlovich et al. reported the observation of $^{174}$Re \cite{1972Ber01}. A mercury target was bombarded with 1 GeV protons from the Leningrad synchrocyclotron. Gamma-ray spectra were measured with a Ge(Li) detector following chemical separation. ``[The table] gives the $\gamma$-lines whose intensities decrease with T$_{l/2}$ = (2.2$\pm$0.2) min. We ascribe this period to the decay of a previously unknown isotope of $^{174}$Re for the following reasons: a) the `accumulation' of $\gamma$-lines of this isotope occurs with a period of $\sim$1~min, which is close to T$_{1/2}$ for $^{174}$Os (45 sec); b) the most intense $\gamma$-rays, 112.4 and 243.6 keV agree well in energy with the transitions 2$^+ \rightarrow$ 0$^+$ and 4$^+ \rightarrow$ 2$^+$ of the rotational band of the ground state of $^{174}$W which are known from the nuclear reactions (111.9 and 243.0 keV).'' This half-life agrees with the currently accepted value of 2.4(4)~min.

\subsection*{$^{175,176}$Re}\vspace{0.0cm}

Nadjakov et al. described the discovery of $^{175}$Re and $^{176}$Re in the 1967 paper ``New isotopes $^{176}$Re and $^{175}$Re'' \cite{1967Nad01}. Holmium and terbium targets were bombarded with beams of $^{16}$O and $^{22}$Ne from the Dubna U-300 heavy-ion accelerator, and $^{175}$Re and $^{176}$Re were produced in 6n and 5n evaporation reactions, respectively. Gamma-ray spectra were measured with a germanium spectrometer following chemical separation. Targets of $^{175}$Ho and $^{159}$Tb were used to synthesize $^{176}$Re and targets of $^{165}$Ho and $^{159}$Tb were used to synthesize $^{175}$Re. ``By varying the ion energy and target thickness, $^{177}$Re could be eliminated so that sources of almost pure $^{176}$Re plus $^{175}$Re could be obtained. The existence of the $^{176}$Re and $^{175}$Re isotopes in our rhenium samples was thus proved.'' For both isotopes a half-life of 5(1)~min was measured. These half-lives agree with the currently accepted values 5.89(5)~min and 5.3(3)~min for $^{175}$Re and $^{176}$Re, respectively.

\subsection*{$^{177,178}$Re}\vspace{0.0cm}

$^{177}$Re and $^{178}$Re were discovered by Haldar and Wiig as reported in the 1957 paper ``New neutron-deficient isotopes of rhenium'' \cite{1957Hal01}. Rhenium targets were bombarded with 120$-$240~MeV protons from the Rochester 130-in synchrocyclotron. Decay curves and positron spectra were measured with beta-proportional and scintillation counters following chemical separation. ``Three new activities have been observed in rhenium obtained by bombardment of rhenium and of tungsten with protons of energies from 40 to 240 Mev and of enriched W$^{180}$ with 10-Mev protons. Positron-emitting Re$^{177}$ of 17-minute half-life was identified through its daughter, the known 2.2-hr W$^{177}$. Evidence is presented for the assignment of Re$^{178}$ to a 15-min, 3.1-Mev positron activity and of Re$^{180}$ to a (20$\pm$1)-hour, 1.9-Mev positron activity.'' These half-lives are consistent with the presently adopted values of 14(1)~min and 13.2(2)~min for $^{177}$Re and  $^{178}$Re, respectively.

\subsection*{$^{179}$Re}\vspace{0.0cm}

Harmatz and Handley reported the discovery of $^{179}$Re in their 1960 paper ``Nuclear spectroscopy of neutron-deficient Lu, Ta, and Re isotopes'' \cite{1960Har01}. Enriched $^{180}$W targets were irradiated with 22 MeV protons from the ORNL 86-in. cyclotron. Conversion electron spectra were measured following chemical separation. ``Targets enriched in W$^{180}$ (6.6\%) were irradiated with protons, and Re was extracted by distillation. An activity attributable to Re$^{179}$ was observed, and a very approximate value of 20$\pm$5 minutes for the half-life was estimated from the rate of decrease of intensity of the conversion lines on successive films.'' This half-life is consistent with the currently accepted value of 19.5(1)~min.

\subsection*{$^{180}$Re}\vspace{0.0cm}

In 1955, Kistiakowsky announced the discovery of $^{180}$Re in the paper ``Metastable states of Re$^{180}$, Ir$^{191}$, Au$^{192}$, Pb$^{201}$, and Pb$^{203}$'' \cite{1955Kis01}. The Berkeley linear accelerator was used to bombard metallic tungsten targets with 31.5~MeV protons. Decay curves and $\gamma$-ray spectra were measured with a gas counter and a scintillation counter. ``[The figure] shows the excitation function obtained for the 145-second activity produced in wolfram. The order of magnitude of the maximum value and the shape of the curve assign the activity to the product of a (p,3n) reaction alone. Thus it must be Re$^{180}$ from W$^{182}$(p,3n)Re$^{180}$.'' This half-life which was assigned to an isomeric state of $^{180}$Re, was included in the calculation of the currently accepted average value of 2.44(6)~min for the $^{180}$Re ground state.

\subsection*{$^{181}$Re}\vspace{0.0cm}

Gallagher et al. reported the discovery of $^{181}$Re in their 1957 paper ``New 20-hour electron-capturing rhenium isotope, Re$^{181}$'' \cite{1957Gal01}. Alpha-particles accelerated to 48~MeV by the Berkeley 60-in. cyclotron bombarded tantalum foils. Beta-rays were measured in a double-focusing spectrometer following chemical separating. ``A new 20$\pm$2 hour electron-capturing rhenium isotope has been investigated. A mass assignment to Re$^{181}$ is made from the energy threshold for its production by alpha-particle bombardment of Ta$^{181}$ and by the chemical separation and identification of its radioactive daughter, W$^{181}$.'' This half-life agrees with the presently accepted value of 19.9(7)~h.

\subsection*{$^{182}$Re}\vspace{0.0cm}

``Neutron deficient radioactive isotopes of rhenium'' was published in 1950 by Wilkinson and Hicks, reporting the discovery of $^{182}$Re \cite{1950Wil02}. Tantalum targets were bombarded with 38~MeV $\alpha$-particles from the Berkeley 60-in. cyclotron. Decay curves and absorption spectra were measured following chemical separation. ``Four new rhenium activities of half-lives 12.7 hours, 64.0 hours, ~240 days, and 2.2 days have been produced by $\alpha$-particle bombardment of tantalum and have been allocated respectively to masses 182, 182, 183, and 184.'' The half-lives of 64.0(5)~h and 12.7(2)~h assigned to $^{182}$Re correspond to the currently adopted values of 64.0(5)~h and 12.7(2)~h for the ground and isomeric state.

\subsection*{$^{183}$Re}\vspace{0.0cm}

In 1950, the observation of $^{183}$Re was reported in the paper ``Os$^{182}$ and Os$^{183}$, new radioactive osmium isotopes'' by Stover \cite{1950Sto01}. At Berkeley, metallic rhenium targets were bombarded with 25~MeV protons from the linear accelerator to produce $^{183}$Re. Magnetic counter and absorption data were taken following chemical separation. ``Bombardment of rhenium (Re$^{186}$, 37.07 percent; Re$^{187}$, 62.93 percent) with 25-Mev protons in the linear accelerator produced the known 97-day Os$^{186}$ and a 12.0-hr. osmium activity which was shown to be the parent of the 120-day Re$^{183}$.'' This half-life is within a factor of two of the currently adopted value of 70.0(14)~d. Previously, Wilkinson and Hicks had reported an approximate half-life of 240~d \cite{1950Wil02}.

\subsection*{$^{184}$Re}\vspace{0.0cm}

In 1940, Fajans and Sullivan observed $^{184}$Re as reported in ``Induced radioactivity of rhenium and tungsten'' \cite{1940Faj01}. Tungsten targets were bombarded with protons and resulting activities and $\gamma$-ray spectra were measured following chemical separation. ``The 52-day activity, which was identified as a rhenium isotope through a series of chemical reactions, has been assigned to mass number 184 for the following reasons. It is produced from rhenium by fast neutrons (n,2n reaction from Re$^{185}$), and appears in rhenium, chemically separated from tungsten bombarded with deuterons (probably a d,n reaction from W$^{183}$, although a d,2n reaction from W$^{184}$ is an additional possibility).'' The  reported 52(2)~d half-life is within a factor of two of the presently accepted value of 38.0(5)~d. The discrepancy is probably due to contaminations from the long-lived isomeric state which was not known until later \cite{2008ENS01}.

\subsection*{$^{185}$Re}\vspace{0.0cm}

Aston reported the discovery of $^{185}$Re in the 1931 article: ``Constitution of rhenium'' \cite{1931Ast04}. $^{185}$Re was observed by chemical volatilization of a sample of rhenium in the Cavendish mass spectrometer. ``Rhenium consists of two isotopes, 185, 187, as was expected from the general rule that complex elements of odd atomic number (above 9) consist of two odd mass numbers two units apart, but it is the first element analysed in which the heavier isotope is the more abundant.''

\subsection*{$^{186}$Re}\vspace{0.0cm}

Sinma and Yamasaki identified $^{186}$Re in the 1939 article ``$\beta$-radioactivities of rhenium'' \cite{1939Sin01}. Metallic rhenium samples were irradiated with slow and fast neutrons produced from Be+D and Li+D reactions from the Tokyo cyclotron, respectively. Energy spectra were measured with a Wilson cloud chamber and decay curves were recorded. ``Now rhenium has only two isotopes Re$^{185}$ and Re$^{187}$. From the above large change in the ratio of the intensities for two cases, it seems therefore more probable, contrary to Pool, Cork and Thornton, to ascribe the 16-hour period to Re$^{188}$ and the 90-hour activity to Re$^{186}$.'' These values agree with the currently accepted values of 3.7183(11)~d and 17.004(22)~h and for $^{186}$Re and $^{188}$Re, respectively. The opposite assignment mentioned in the quote was published two years earlier \cite{1937Poo01}. Half-lives of 20~h \cite{1935Ama01,1935Kur02} and 85~h \cite{1935Kur02} had previously been reported without mass assignments.

\subsection*{$^{187}$Re}\vspace{0.0cm}

Aston reported the discovery of $^{187}$Re in the 1931 article: ``Constitution of rhenium'' \cite{1931Ast04}. $^{187}$Re was observed by chemical volatilization of a sample of rhenium in the Cavendish mass spectrometer. ``Rhenium consists of two isotopes, 185, 187, as was expected from the general rule that complex elements of odd atomic number (above 9) consist of two odd mass numbers two units apart, but it is the first element analysed in which the heavier isotope is the more abundant.''

\subsection*{$^{188}$Re}\vspace{0.0cm}

Sinma and Yamasaki identified $^{188}$Re in the 1939 article ``$\beta$-radioactivities of rhenium'' \cite{1939Sin01}. Metallic rhenium samples were irradiated with slow and fast neutrons produced from Be+D and Li+D reactions from the Tokyo cyclotron, respectively. Energy spectra were measured with a Wilson cloud chamber and decay curves were recorded. ``Now rhenium has only two isotopes Re$^{185}$ and Re$^{187}$. From the above large change in the ratio of the intensities for two cases, it seems therefore more probable, contrary to Pool, Cork and Thornton, to ascribe the 16-hour period to Re$^{188}$ and the 90-hour activity to Re$^{186}$.'' These values agree with the currently accepted values of 3.7183(11)~d and 17.004(22)~h and for $^{186}$Re and $^{188}$Re, respectively. The opposite assignment mentioned in the quote was published two years earlier \cite{1937Poo01}. Half-lives of 20~h \cite{1935Ama01,1935Kur02} and 85~h \cite{1935Kur02} had previously been reported without mass assignments.

\subsection*{$^{189}$Re}\vspace{0.0cm}

In 1963, Crasemann et al. observed $^{189}$Re as described in the paper ``Properties of radioactive Re$^{189}$'' \cite{1963Cra01}. Metallic osmium was irradiated with neutrons produced by bombarding beryllium with 20 MeV deuterons from the Brookhaven 60-in. cyclotron and $^{189}$Re was produced in (n,p) and (n,pn) reactions on $^{189}$Os and $^{190}$Os, respectively. In addition, the reaction $^{185}$W($\alpha$,p)$^{189}$Re was studied. Gamma-ray and conversion electron spectra were measured following chemical separation. ``The half-life was determined by integrating areas under the 217- and 219-keV gamma-ray peaks in scintillation spectra that were recorded at intervals over periods of approximately five days each, using sources from three different bombardments (one Os+n, two W+$\alpha$). The result obtained for the half-life of Re$^{189}$ is 23.4$\pm$1.0~h.''  This half-life agrees with the presently adopted value of 24.3(4)~h. Previously reported half-lives of $\sim$150~d or $<$5~y \cite{1951Lin02}, 250$-$300~d \cite{1951Tur02}, and 120~d \cite{1962Bli01} were evidently incorrect. Although Flegenheimer et al. \cite{1963Fle01} submitted their results of a 23~h half-life nine days earlier we still credit Crasemann with the discovery because Flegenheimer et al. specifically acknowledged the work by Crasemann.

\subsection*{$^{190}$Re}\vspace{0.0cm}

Aten and de Feyfer reported the discovery of $^{190}$Re in the 1955 paper ``Rhenium 190'' \cite{1955Ate03}. Osmium targets were irradiated with 26 MeV deuterons and fast neutrons from the Philips' synchro-cyclotron. Decay curves as well as absorption- and $\gamma$-ray spectra were measured following chemical separation. ``The fact that the 2.8-minutes rhenium is formed both by neutron and by deuton irradiation, suggests that it may well be $^{190}$Re, which can be formed by the reactions $^{180}$Os(n,p) and $^{192}$Os(d,$\alpha$).'' The reported half-life of 2.8(5)~min agrees with the currently accepted value of 3.1(3)~min.

\subsection*{$^{191}$Re}\vspace{0.0cm}

In 1963, Crasemann et al. identified $^{191}$Re in the paper ``Properties of radioactive Re$^{189}$'' \cite{1963Cra01}. During the study of $^{189}$Re by (n,p) and (n,pn) reactions on metallic osmium irradiated with neutrons produced by bombarding beryllium with 20 MeV deuterons from the Brookhaven 60-in. cyclotron the previously observed 9.75~min half-life \cite{1953Ate01} was assigned to $^{191}$Re. ``Aten and de Feyfer obtained a 9.75-min rhenium activity by bombardment of
osmium with fast neutrons from 26-MeV deuterons on brass, and assigned this half-life to mass number 189, 190, or 192. It is now clear that they produced 10-min
Re$^{191}$ through the reaction Os$^{192}$(n,pn).'' We credit Craseman et al. with the discovery because apparently no further results for the half-life of $^{191}$Re were published in the refereed literature.

\subsection*{$^{192}$Re}\vspace{0.0cm}

In ``D\'esint\'egration du rh\'enium 192'' Blachot et al. reported the observation of $^{192}$Rh in 1965 \cite{1965Bla01}. An enriched $^{192}$Os target was irradiated with 14$-$15~MeV neutrons produced at the Grenoble 400~kV accelerator. $^{192}$Re was identified by measuring $\gamma$-ray spectra with a NaI(Tl) scintillator. ``\'Etude par spectrom\'etrie $\gamma$ du rh\'enium 192 produit par la r\'eaction nucl\'eaire $^{192}$Os(n,p)$^{192}$Re avec des neutrons de 14$-$15 MeV. Le spectre $\gamma$ de d\'esexcitation de $^{192}$Os a \'et\'e mis en \'evidence ainsi que la p\'eriode T$_{1/2} \sim$ 6.2$\pm$0.8 s du $^{192}$Re.'' [$^{192}$Re produced in the nuclear reaction $^{192}$Os(n,p)$^{192}$Re with 14$-$15 MeV neutrons was studied by $\gamma$-spectroscopy. A half-life of T$_{1/2} \sim$ 6.2$\pm$0.8~s for $^{192}$Re was identified by the $\gamma$-spectrum of the $^{192}$Os deexcitation.] This half-life is not mentioned in the ENSDF database which instead lists a 16(6)~s referring to a private communication \cite{2008ENS01}.

\subsection*{$^{193,194}$Re}\vspace{0.0cm}

Benlliure et al. published the discovery of $^{193}$Re and $^{194}$Re in the 1999 paper entitled ``Production of neutron-rich isotopes by cold fragmentation in the reaction $^{197}$Au + Be at 950 \emph{A} MeV'' \cite{1999Ben01}. A 950~A$\cdot$MeV $^{197}$Au beam from the SIS synchrotron of GSI was incident on a beryllium target and $^{193}$Re and $^{194}$Re were produced in projectile fragmentation reactions. The FRS fragment separator was used to select isotopes with a specific mass-to-charge ratio.  ``In the right part of [the figure] the projected A/Z distributions are shown for the different elements transmitted in this setting of the FRS. In this setting the isotopes $^{193}$Re, $^{194}$Re, $^{191}$W, $^{192}$W, $^{189}$Ta, $^{187}$Hf and $^{188}$Hf were clearly identified for the first time. Only isotopes with a yield higher than 15 counts were considered as unambiguously identified.''

\subsection*{$^{195,196}$Re}\vspace{0.0cm}

The first refereed publication of the observation of $^{195}$Re and $^{196}$Re was the 2008 paper ``Single-particle behavior at N = 126: Isomeric decays in neutron-rich $^{204}$Pt'' by Steer et al. \cite{2008Ste01}. A 1 GeV/A $^{208}$Pb beam from the SIS-18 accelerator at GSI impinged on a $^9$Be target and the projectile fragments were selected and identified in-flight by the Fragment Separator FRS. The observation of the new neutron-rich rhenium isotopes was not specifically mentioned but $^{195}$Re and $^{196}$Re events are clearly visible and identified in the particle identification plot in the first figure.

\section{Discovery of $^{161-199}$Os}

Thirty-nine osmium isotopes from A = 161--199 have been discovered so far; these include 7 stable, 24 neutron-deficient and 8 neutron-rich isotopes.
According to the HFB-14 model \cite{2007Gor01}, $^{257}$Os should be the last odd-even particle stable neutron-rich nucleus while the even-even particle stable neutron-rich nuclei should continue through $^{260}$Os. At the proton dripline three more particle stable osmium isotopes are predicted ($^{158-160}$Os) and in addition seven more isotopes could possibly still have half-lives longer than 10$^{-9}$~s \cite{2004Tho01}. Thus, about 70 isotopes have yet to be discovered corresponding to 64\% of all possible osmium isotopes.

In 2004, J.W. Arblaster published a review article entitled ``The Discoverers of the Osmium Isotopes'' \cite{2004Arb01}. Although he selected slightly different criteria for the discovery, our assignments agree in most of the cases. Since then five additional isotopes ($^{161}$Os, $^{195}$Os, and $^{197-199}$Os) were discovered.

\begin{figure}
	\centering
	\includegraphics[scale=.7]{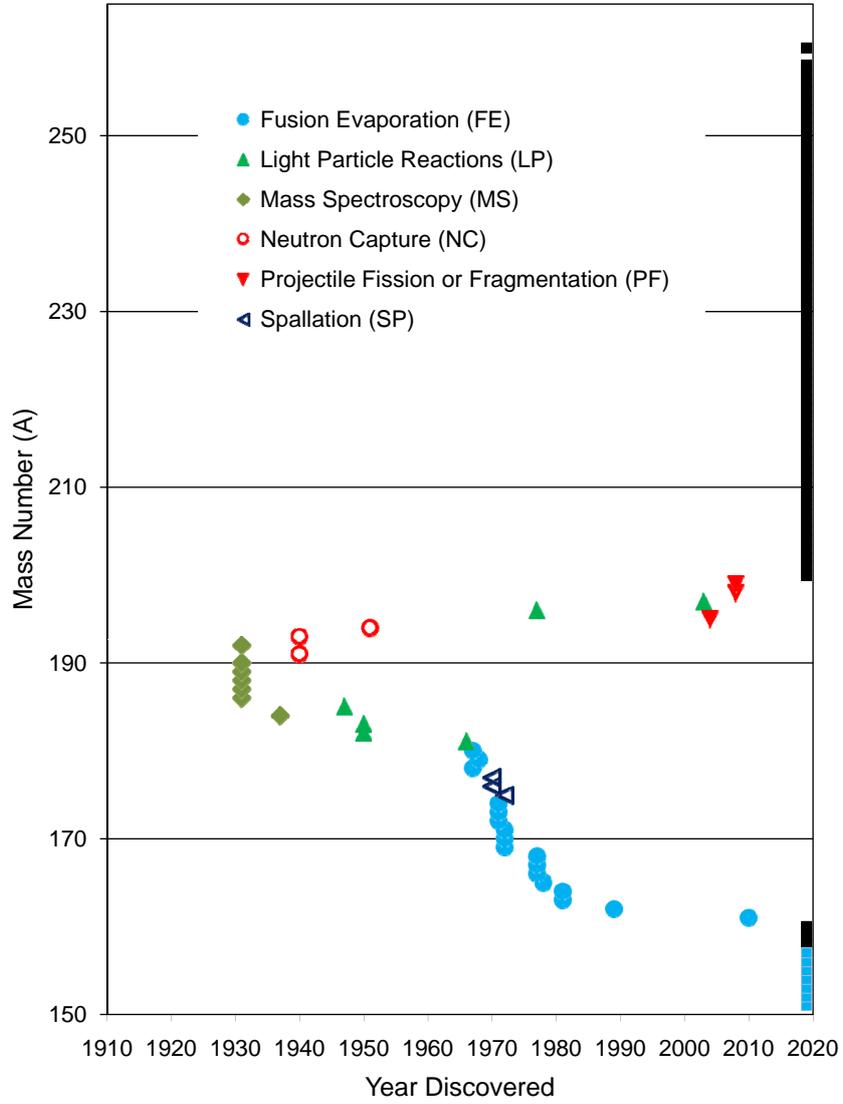}
	\caption{Osmium isotopes as a function of time when they were discovered. The different production methods are indicated. The solid black squares on the right hand side of the plot are isotopes predicted to be bound by the HFB-14 model. On the proton-rich side the light blue squares correspond to unbound isotopes predicted to have half-lives larger than $\sim 10^{-9}$~s.}
\label{f:year-os}
\end{figure}

Figure \ref{f:year-os} summarizes the year of first discovery for all osmium isotopes identified by the method of discovery. The range of isotopes predicted to exist is indicated on the right side of the figure. The radioactive osmium isotopes were produced using fusion evaporation reactions (FE), light-particle reactions (LP), neutron capture reactions (NC), spallation reactions (SP), and projectile fission or fragmentation (PF). The stable isotopes were identified using mass spectroscopy (MS). Light particles also include neutrons produced by accelerators. In the following, the discovery of each osmium isotope is discussed in detail.

\subsection*{$^{161}$Os}\vspace{0.0cm}

In 2010, Bianco et al. discovered $^{161}$Os, and reported their findings in the paper ``Discovery of $^{157}$W and $^{161}$Os'' \cite{2010Bia01}. At the University of Jyv\"askyl\"a in Finland, self-supporting $^{106}$Cd targets were bombarded with 290, 300, and 310~MeV $^{58}$Ni beams. $^{161}$Os was produced in the fusion-evaporation reaction $^{106}$Cd($^{58}$Ni,3n) and identified with the GREAT spectrometer where $\alpha$-spectra were measured following the RITU gas-filled separator. ``The clear peak comprising ~200 counts at 6890$\pm$12 keV is a new activity that we assign as the $\alpha$ decay of $^{161}$Os.'' The measured half-life of 640(60)~$\mu$s corresponds to the currently adopted value.

\subsection*{$^{162}$Os}\vspace{0.0cm}

In the 1989 paper ``The new nuclei $^{162}$Os and $^{156}$Ta and the N=84 alpha emitting isomers'' by Hofmann et al. the discovery of $^{162}$Os was announced \cite{1989Hof01}. Enriched $^{106}$Cd targets were irradiated with 47$-$89~MeV $^{58}$Ni beams from the GSI UNILAC. $^{162}$Os was produced in the fusion-evaporation reaction $^{106}$Cd($^{58}$Ni,2n) identified with the velocity filter SHIP. ``In the reaction $^{58}$Ni + $^{106}$Cd $\rightarrow ^{164}$Os at E=47.2~MeV we observed a new $\alpha$ line of (6611$\pm$30) keV energy correlated to the $^{158}$W line. We assign this line to the $\alpha$ decay of the new isotope $^{162}$Os.'' The measured half-life of 1.9(7)~ms agrees with the currently accepted value of 2.1(1)~ms.

\subsection*{$^{163,164}$Os}\vspace{0.0cm}

The 1981 paper ``New neutron deficient isotopes in the range of elements Tm to Pt'' reported the discovery of $^{163}$Os and $^{164}$Os by Hofmann et al. at GSI using the linear accelerator UNILAC. \cite{1981Hof01}. $^{163}$Os and $^{164}$Os  were produced in reactions bombarding neutron deficient targets between molybdenum and tin with a beam of $^{58}$Ni as well as bombarding targets between vanadium and nickel with a beam of $^{107}$Ag, with energies between 4.4 MeV/u and 5.9 MeV/u. Residues were separated using the velocity filter SHIP. For $^{163}$Os no half-life was measured and only an $\alpha$-decay energy of 6510(30)~keV was listed in a table. ``The lightest isotope, $^{168}$Pt, could be identified by 4 correlated events to the daughter $^{164}$Os. This again is a new isotope, clearly identified in correlations to its daughter and granddaughter, $^{160}$W and $^{156}$Hf, respectively, as can be seen in [the figure].'' The measured half-life of 41(20)~ms agrees with the presently accepted value of 21(1)~ms.

\subsection*{$^{165}$Os}\vspace{0.0cm}

``Copper ion induced reactions on $^{110-108-106}$Cd, $^{109-107}$Ag and $^{110}$Pd. New rhenium, osmium and iridium isotopes'' was published in 1978 by Cabot et al. announcing the discovery of $^{165}$Os \cite{1978Cab01}. A 400~MeV $^{63}$Cu beam from the ALICE accelerator at Orsay, France, bombarded isotopically enriched $^{106}$Cd target to populate $^{165}$Os in the fusion-evaporation reaction $^{106}$Cd($^{63}$Cu,p3n). Alpha particles from fragments collected by a He-jet were detected to determine the decay energies and half-lives. ``From [the figure] it is clear that we have only the rising part of the excitation function for the reaction emitting one extra particle and we propose to assign this 6.20~MeV $\alpha$-ray to $^{165}$Os formed by the (Cu,p3n) reaction. The $\alpha$ systematics are again consistent with those mass assignments.''

\subsection*{$^{166-168}$Os}\vspace{0.0cm}

Cabot et al. published the first observation of $^{166}$Os, $^{167}$Os, and $^{168}$Os in the paper ``New osmium and iridium isotopes produced through $^{63}$Cu induced reactions'' in 1977 \cite{1977Cab01}. Self-supporting $^{106}$Cd and $^{109}$Ag targets were bombarded with a 380~MeV $^{63}$Cu beam from Orsay ALICE accelerator. The neutron-deficient osmium isotopes were produced in the fusion-evaporation reactions $^{106}$Cd($^{63}$Cu,p2n)$^{166}$Os, $^{109}$Ag($^{63}$Cu,5n)$^{167}$Os, and $^{109}$Ag($^{63}$Cu,4n)$^{168}$Os. Reaction fragments were collected with a He-jet and $\alpha$-decay half-lives and decay energies were measured.  ``New $\alpha$ active osmium and iridium isotopes $^{168}$Os, $^{167}$Os, $^{166}$Os, and $^{170}$Ir have been identified by cross bombardments and excitation functions measurements.'' The measured half-lives of 0.3(1)~s ($^{166}$Os), 0.65(15)~s ($^{167}$Os), and  2.0(4)~s ($^{168}$Os) agree with the presently adopted value of 216(9)~ms, 810(60)~ms, and 2.06(6)~s, respectively.

\subsection*{$^{169}$Os}\vspace{0.0cm}

``Evidence for the $\alpha$ decay of the new isotope $^{169}$Os'' reported the discovery of $^{169}$Os in 1972 by Toth et al. \cite{1972Tot01}. Enriched $^{156}$Dy targets were bombarded with a $^{20}$Ne beam from the Oak Ridge isochronous cyclotron. $^{169}$Os was produced in the fusion-evaporation reaction $^{156}$Dy($^{20}$Ne,7n) and identified by measuring $\alpha$-decay spectra. ``In a series of $^{20}$Ne bombardments of $^{156}$Dy a new $\alpha$ emitter was identified with a half-life of 3.0$\pm$0.5 sec and an $\alpha$-decay energy of 5.56$\pm$0.02~MeV. On the basis of $\alpha$-decay systematics and the variation of its yield with bombarding energy, the most likely nuclidic assignment for this $\alpha$ group is $^{169}$Os.'' This half-life agrees with the currently adopted value of 3.46(11)~s.

\subsection*{$^{170,171}$Os}\vspace{0.0cm}

In the 1972 paper ``Alpha-decay properties of the new osmium isotopes, $^{170}$Os and $^{171}$Os'' Toth et al. described the discovery of $^{170}$Os and $^{171}$Os \cite{1972Tot02}. Enriched $^{156}$Dy targets were bombarded with a $^{20}$Ne beam with energies up to 160~MeV from the Oak Ridge isochronous cyclotron and $^{170}$Os and $^{171}$Os were identified by measuring $\alpha$-decay spectra. ``It is seen that not only do our results support the assignment of the 5.105-MeV $\alpha$ group to $^{172}$Os, but the ($^{20}$Ne,5n) and ($^{20}$Ne,6n) curves reproduce the data points for the new $\alpha$ activities reasonably well. The indication then is that the 5.23- and 5.40-MeV $\alpha$ group are due to the decay of $^{171}$Os and $^{170}$Os, respectively.'' The measured half-lives of 7.1(5)~s ($^{170}$Os) and 8.2(8)~s $^{171}$Os agree with the currently adopted values of 7.37(18)~s and 8.3(2)~s.

\subsection*{$^{172-174}$Os}\vspace{0.0cm}

The discovery of $^{172}$Os, $^{173}$Os, and $^{174}$Os was reported in 1971 in ``Alpha decay of neutron-deficient osmium isotopes'' by Borggreen and Hyde \cite{1971Bor01}. The Berkeley heavy-ion linear accelerator accelerated $^{16}$O to 110$-$160~MeV and bombarded enriched $^{164}$Er and $^{166}$Er. The reaction products were positioned in front of a semiconducting silicon detector by a helium-jet transport system. ``Three neutron-deficient isotopes of osmium have been produced by the interaction of $^{16}$O ions with erbium targets and observed by their $\alpha$-decay. They are $^{172}$Os, E$_\alpha$ = 5.105~MeV, t$_{1/2}$ = 19~s; $^{173}$Os, E$_\alpha$  = 4.94~MeV, t$_{1/2}$ = 16~s; and $^{174}$Os, E$_\alpha$  = 4.76~MeV, t$_{1/2}$ = 45~s.'' These half-lives agree with the presently adopted values of  19.2(9)~s, 16(5)~s, and 44(4)~s.

\subsection*{$^{175}$Os}\vspace{0.0cm}

In 1972, the paper ``Short-lived osmium isotopes'' was published reporting the discovery of $^{175}$Os by Berlovich et al. \cite{1972Ber01}. Mercury nitrate was bombarded with 1~GeV protons from the Leningrad synchrocyclotron. Gamma-ray spectra were measured with a Ge(Li)-detector following chemical separation.  The mean half-life value assigned in this experiment to $^{175}$Os was 1.4$\pm$0.1 min, which agrees with the currently accepted value. ``There is, however, some uncertainty: we are observing either the decay of a previously unknown isotope $^{175}$Os or the decay of an unknown isomer of one of the known isotopes of osmium. The first alternative appears to us to be the most probable for the following reasons. First, in the daughter products of the decay of our samples we observed known $\gamma$-lines of $^{175}$Ta (126, 248, and 267~keV). Thus, $^{175}$Os is present in the samples we studied.'' The reported half-life of 1.5~min agrees with the presently adopted value of 1.4(1)~min.

\subsection*{$^{176,177}$Os}\vspace{0.0cm}

In 1970, Arlt et al. discovered $^{176}$Os and $^{177}$Os as reported in their paper ``New osmium isotopes $^{176}$Os and $^{177}$Os, decay of $^{177-180}$Re and $^{178-180}$Os, and decay scheme of $^{179}$Re'' \cite{1970Arl01}. The Dubna JINR synchrocyclotron accelerated protons to 660~MeV which bombarded metallic gold targets and $\gamma$ spectra were measured following chemical separation. ``Our results for the half-lives of the new $^{176}$Os and $^{177}$Os isotopes are given in [the figure]. For these measurements we used $\gamma$ lines of the $^{176}$W, $^{177}$W and $^{176}$Ta descendants. The half-lives of the new $^{176}$Os and $^{177}$Os isotopes are 3.0$\pm$0.7 min and 3.5$\pm$0.8 min, respectively.'' These values agree with the accepted values of 3.6(5)~min and  3(2)~min, respectively.

\subsection*{$^{178}$Os}\vspace{0.0cm}

``Ground state (quasi-) rotational levels in light Os, Pt and Hg nuclei'', by Burde et al., reported the first observation of $^{178}$Os in 1967 \cite{1967Bur01}. A 93~MeV $^{14}$N beam from the Berkeley Hilac bombarded a $^{169}$Tm target and $^{178}$Os was formed in the fusion-evaporation reaction $^{169}$Tm($^{14}$N,5n). Electron and $\gamma$-ray spectra were measured and the rotational band of $^{178}$Os was observed up to the 12$^+$ state. ``Energy levels in some neutron-deficient doubly even nuclei in the platinum region have been studied following heavy-ion reactions. Information on the ground state rotational (or quasi-rotational) bands in $^{178,180,182}$Os, $^{182,184,186,188}$Pt, and $^{188,190}$Hg is presented.'' A year later Belyaev et al. reported the first half-life measurement of $^{178}$Os \cite{1968Bel01}.

\subsection*{$^{179}$Os}\vspace{0.0cm}

The first identification of $^{179}$Os was published by Belyaev et al. in their 1968 paper ``New osmium isotopes: $^{179}$Os and $^{178}$Os. Identification and gamma spectra of $^{179}$Re, $^{178}$Re, $^{177}$Re, $^{177}$W, $^{180}$W, $^{180}$Os and $^{181}$Os'' \cite{1968Bel01}. Carbon and nitrogen beams from the Dubna U-150 cyclotron at 6.7~MeV/nucleon bombarded ytterbium and thulium targets, respectively. Gamma-ray spectra were recorded with a lithium drifted germanium detector following chemical separation. ``The 750 and 1300~keV $\gamma$ rays (which decay with a half life of 8~min) can be ascribed to decay of $^{179}$Os, and the 120, 230, 290, 430 and 920~keY $\gamma$ rays can be associated with accumulation and decay of the 20~min $^{179}$Re daughter.'' This half-life is consistent with the presently accepted value of 6.5(3)~min.

\subsection*{$^{180}$Os}\vspace{0.0cm}

The discovery of $^{180}$Os was reported in the 1967 paper ``The decay of the isotope Os$^{180}$'' by Belyaev et al. \cite{1967Bel01}. Tu$_2$O$_3$ targets were bombarded with 100~MeV $^{14}$N from the Dubna U-150 cyclotron and $^{180}$Os was formed in the fusion-evaporation reaction $^{169}$Th($^{15}$N,4n). Gamma-ray spectra were measured with a NaI(Tl) crystal following chemical separation. ``In our measurements we observed $\gamma$ lines having the same energy, 105 and 510 keV, and an intense 880-keV $\gamma$ line. This gives grounds for assuming that the observed 21-minute activity is connected with the decay of the isotope Os$^{180}$, which, taking into account the time necessary to separate the osmium ($\sim$1 hour), is in radioactive equilibrium with the daughter rhenium, that is, Os$^{180}\stackrel{21~min}{\longrightarrow}$Re$^{180}\stackrel{2.4~min}{\longrightarrow}$.'' The reported half-life of 21(2)~min agrees with the currently accepted value of 21.5(4)~min. Previously a 23~min half-life had been tentatively assigned to $^{181}$Os \cite{1958Fos01} and was later reassigned to $^{180}$Os \cite{1966Hof01}.

\subsection*{$^{181}$Os}\vspace{0.0cm}

Hofstetter and Daly reported the identification of $^{181}$Os in their 1966 paper ``Decay properties of neutron deficient osmium and rhenium isotopes. I. Decay modes of Re$^{179}$, Os$^{180}$, and Os$^{181}$'' \cite{1966Hof01}. The Argonne 60-in. cyclotron as well as the Oak Ridge 88-in. cyclotron were used to bombard enriched $^{182}$W targets with 32~MeV $^3$He and 65~MeV $^4$He. Gamma-ray spectra were measured with NaI(Tl) crystals and lithium drifted germanium detectors following chemical separation. ``The high quality of the spectral data is indicated in [the figure], which shows clearly the marked changes produced in the low-energy $\gamma$ spectrum as the l05-min osmium activity decays into 20-h Re$^{181}$. The l05-min activity is therefore assigned to Os$^{181}$ with confidence.'' The reported 105(3)~min half-life corresponds to the currently accepted value. A previous tentative assignment of a 23~min half-life to $^{181}$Os by Foster et al. \cite{1958Fos01} was later reassigned to $^{180}$Os \cite{1966Hof01}. Foster also assigned a 2~min half-life to $^{180}$Os which probably corresponds to the isomeric state of $^{181}$Os. The previously observed half-lives of 2.7~h \cite{1960Sur01} and 2.5~h \cite{1966Bel01} have not been considered sufficiently clean to warrant the claim of discovery \cite{1966Bed01,1966Hof01,1969Hug01,2004Arb01}.

\subsection*{$^{182,183}$Os}\vspace{0.0cm}

In 1950, the discovery of $^{182}$Os and $^{183}$Os was announced in the paper ``Os$^{182}$ and Os$^{183}$, new radioactive osmium isotopes'' by Stover \cite{1950Sto01}. At Berkeley, metallic rhenium targets were bombarded with 40~MeV from the 184-in. cyclotron and 25~MeV protons from the linear accelerator to produce $^{182}$Os and $^{183}$Os, respectively. Magnetic counter and absorption data were taken following chemical separation. ``With 40-Mev protons in the 184-in. cyclotron, an additional activity of 24-hr. half-life was formed which decayed to the 12.7-hr. Re$^{182}$. The 24-hr. Os$^{182}$ decays by electron capture, no positrons having been detected... Os$^{183}$ decays by electron capture, and emits conversion electrons of energies 0.15~Mev and 0.42~Mev, and gamma-rays of energies 0.34~Mev and 1.6~Mev.'' The measured half lives of 24(1)~h ($^{182}$Os) and 12.0(5)~h ($^{182}$Os) agree with the currently accepted values of 22.10(25)~h and 13.0(5)~h, respectively.

\subsection*{$^{184}$Os}\vspace{0.0cm}

In 1937, Nier published the discovery of $^{184}$Os in his paper ``The isotopic constitution of osmium'' \cite{1937Nie01}. OsO$_{4}$ vapor was fed to a high resolving, high intensity mass spectrometer designed for the detection of rare isotopes. ``It is the purpose of this communication to show that in addition to the above isotopes an extremely rare isotope, Os$^{184}$, exists, present to about one part in 5700 in osmium.''

\subsection*{$^{185}$Os}\vspace{0.0cm}

In the paper ``Radioactive isotopes of Re, Os, and Ir'', Goodman and Pool described their discovery of $^{185}$Os in 1947 \cite{1947Goo01}. $^{185}$Os was produced in (d,2n) reactions on rhenium. Decay curves were measured following chemical separation. ``A new period has been found in the osmium
fraction after a deuteron bombardment of Re. The chemical separations were carried out as previously described. The decay of this isotope
is shown in [the figure] and is seen to have a half-life of 94.7$\pm$2.0 days. The isotope is predominately $\gamma$-ray emitting and has been tentatively placed at Os$^{185}$.'' This half life agrees with the accepted value of 93.6(5)~d.

\subsection*{$^{186-190}$Os}\vspace{0.0cm}

In 1931, Aston reported the first observation of the stable osmium isotopes $^{186}$Os, $^{187}$Os, $^{188}$Os, $^{189}$Os, and $^{190}$Os in ``Constitution of osmium and ruthenium'' \cite{1931Ast03}. Osmium tetroxide was used in the Cavendish mass spectrograph. ``In consequence the admission could only be by very small periodical doses during the exposure, and it was only with the greatest difficulty that spectra of adequate density were obtained. These indicate four strong isotopes and two very weak ones, one of the latter being isobaric with tungsten, W$^{186}$. Fortunately it was easy to photograph on the same plate several short exposures of the mercury group, which is sufficiently near in mass to provide a reasonably reliable density scale. The mass numbers and provisional relative abundance are as follows: Mass-number (Percentage abundance): 186 (1.0), 187 (0.6), 188 (13.5), 189 (17.3), 190 (25.1), 192 (42.6).''

\subsection*{$^{191}$Os}\vspace{0.0cm}

Zingg reported the observation of $^{191}$Os in the 1940 paper ``Die Isobarenpaare Cd-In, In-Sn, Sb-Te, Re-Os'' \cite{1940Zin01}. Neutrons from a Ra-Be source irradiated osmium targets and X-rays were measured. ``Weil beim Osmium die Isotopen der Massenzahlen A = 186, 187, 188, 189, 190 und 192 existieren, k\"onnen durch langsame Neutronen nur die instabilen Kerne Os$^{191}_{76}$ und Os$^{193}_{76}$ entstehen, und weil Os$^{192}_{76}$ das h\"aufigste Isotop ist, wird man folgende Zuordnung treffen: T= 30~h Os$^{192}_{76}$ + n$^1_0 \rightarrow$ Os$^{193}_{76} \rightarrow$ Ir$^{193}_{76}$ + $e^-$, T = 10~d Os$^{190}_{76}$ + n$^1_0 \rightarrow$ Os$^{191}_{76} \rightarrow$ Ir$^{191}_{76}$ + $e^-$.'' [Because the existing osmium isotopes have mass numbers 186, 187, 188, 189, 190, and 192, only the unstable nuclei Os$^{191}_{76}$ and Os$^{193}_{76}$ will be produced and because Os$^{192}_{76}$ is the most abundant isotope, the following assignment is made: T= 30~h Os$^{192}_{76}$ + n$^1_0 \rightarrow$ Os$^{193}_{76} \rightarrow$ Ir$^{193}_{76}$ + $e^-$, T = 10~d Os$^{190}_{76}$ + n$^1_0 \rightarrow$ Os$^{191}_{76} \rightarrow$ Ir$^{191}_{76}$ + $e^-$.] The 10~d half-life is consistent with the currently adopted value 15.4(1)~d.

\subsection*{$^{192}$Os}\vspace{0.0cm}

In 1931, Aston reported the first observation of stable $^{192}$Os in ``Constitution of osmium and ruthenium'' \cite{1931Ast03}. Osmium tetroxide was used in the Cavendish mass spectrograph. ``In consequence the admission could only be by very small periodical doses during the exposure, and it was only with the greatest difficulty that spectra of adequate density were obtained. These indicate four strong isotopes and two very weak ones, one of the latter being isobaric with tungsten, W$^{186}$. Fortunately it was easy to photograph on the same plate several short exposures of the mercury group, which is sufficiently near in mass to provide a reasonably reliable density scale. The mass numbers and provisional relative abundance are as follows: Mass-number (Percentage abundance): 186 (1.0), 187 (0.6), 188 (13.5), 189 (17.3), 190 (25.1), 192 (42.6).''

\subsection*{$^{193}$Os}\vspace{0.0cm}

Zingg reported the observation of $^{193}$Os in the 1940 paper ``Die Isobarenpaare Cd-In, In-Sn, Sb-Te, Re-Os'' \cite{1940Zin01}. Neutrons from a Ra-Be source irradiated osmium targets and X-rays were measured. ``Weil beim Osmium die Isotopen der Massenzahlen A = 186, 187, 188, 189, 190 und 192 existieren, k\"onnen durch langsame Neutronen nur die instabilen Kerne Os$^{191}_{76}$ und Os$^{193}_{76}$ entstehen, und weil Os$^{192}_{76}$ das h\"aufigste Isotop ist, wird man folgende Zuordnung treffen: T= 30~h Os$^{192}_{76}$ + n$^1_0 \rightarrow$ Os$^{193}_{76} \rightarrow$ Ir$^{193}_{76}$ + $e^-$, T = 10~d Os$^{190}_{76}$ + n$^1_0 \rightarrow$ Os$^{191}_{76} \rightarrow$ Ir$^{191}_{76}$ + $e^-$.'' [Because the existing osmium isotopes have mass numbers 186, 187, 188, 189, 190, and 192, only the unstable nuclei Os$^{191}_{76}$ and Os$^{193}_{76}$ will be produced and because Os$^{192}_{76}$ is the most abundant isotope, the following assignment is made: T= 30~h Os$^{192}_{76}$ + n$^1_0 \rightarrow$ Os$^{193}_{76} \rightarrow$ Ir$^{193}_{76}$ + $e^-$, T = 10~d Os$^{190}_{76}$ + n$^1_0 \rightarrow$ Os$^{191}_{76} \rightarrow$ Ir$^{191}_{76}$ + $e^-$.] The 30~h half-life is consistent with the currently adopted value 30.11(1)~h. In 1935, Kurtchatow et al. reported an osmium half-life of 40~h without a mass assignment \cite{1935Kur01}.

\subsection*{$^{194}$Os}\vspace{0.0cm}

In 1951, Lindner reported the first observation of $^{194}$Os in the paper ``Characteristics of some radionuclides of tungsten, rhenium, and osmium formed by second-order thermal neutron capture'' \cite{1951Lin02}. Osmium targets were irradiated at Oak Ridge National Laboratory and chemically separated and the activity counted after several months. ``Since the principal long-lived activity present in thermal-neutron activated osmium is the 97-day Os$^{186}$, the difficulty encountered in observing a second-order product such as Os$^{194}$ is similar to that described for tungsten. Again, however, because Os$^{194}$ would necessarily decay to the known 19-hour Ir$^{194}$, the elucidation of the parent through its radioactive daughter seemed the most feasible approach... An accurate value for the half-life of the Os$^{194}$ has not thus far been feasible by this method because its very long half-life still renders errors in mounting and counting from sample to sample appreciable as compared with the fraction decayed. However, the half-life appears to be around 700 days. Since direct decay measurements of the osmium itself indicate that the shorter-lived Os$^{185}$ is gradually giving way to Os$^{194}$, it will be possible within two years to observe the latter directly.'' Although the half-life is significantly shorter than the presently accepted value of 6.0(2)~y we credit Lindner with the discovery because of the correct identification of the daughter activity.

\subsection*{$^{195}$Os}\vspace{0.0cm}

``$^{136}$Ba studied via deep-inelastic collisions: Identification of the SYNTAX isomer'' was published in 2004 reporting the observation of $^{195}$Os by Valiente-Dob\'on et al. \cite{2004Val01}. A self-supporting enriched $^{198}$Pt target was bombarded with an 850~MeV $^{136}$Xe from the Berkeley 88-in. cyclotron and $\gamma$ rays were measured with GAMMASPHERE in coincidence with recoil products in the parallel plate avalanche counter Chico. ``A wide range of isomeric states with half-lives in the nanosecond-to-microsecond range were populated in both the xenon and platinum regions...  Note that the isomers found in $^{131}$I, $^{133}$I, $^{184}$W, $^{191}$Os, $^{192}$Os, and $^{198}$Pt have not been reported in the literature prior to the current work.'' This represents the first observation of $^{195}$Os because a previously reported 6.5~min half-life \cite{1957Bar01} was later reassigned to $^{81}$Rb \cite{1974Col01,1999Zho01}.

\subsection*{$^{196}$Os}\vspace{0.0cm}

Haustein et al. reported the discovery of $^{196}$Os in the 1977 paper ``New neutron-rich isotope: $^{196}$Os'' \cite{1977Hau01}. Isotopically enriched $^{198}$PtCl$_{4}$ targets were irradiated with neutrons at the Brookhaven Medium Energy Intense Neutron (MEIN) Facility. X-, $\gamma$-, and $\beta$ rays were measured following chemical separation. ``While the observation of $^{196}$Ir in secular equilibrium provides the strongest evidence for correct assignment of the new radioactivity to $^{196}$Os, corroborative evidence of this assignment was obtained from x-ray spectra.'' The reported half-life of 34.9(4)~min corresponds to the currently accepted value.

\subsection*{$^{197}$Os}\vspace{0.0cm}

The discovery of $^{197}$Os was reported in the 2003 paper ``Observation of $^{197}$Os'' by Xu et al. \cite{2003Xu01}. Natural platinum foils were irradiated with 14-MeV neutrons. $^{197}$Os was produced in the reaction $^{198}$Pt(n,2p) and identified by measuring $\gamma$- and X-rays. ``The ten new $\gamma$-rays of 41.2, 50.7, 196.8, 199.6, 223.9, 233.1, 250.2, 342.1, 403.6, and 406.4~keV assigned to the decay of $^{197}$Os were observed. The half-life of $^{197}$Os has been determined as 2.8$\pm$0.6 minutes.'' This half-life corresponds to the presently adopted value.

\subsection*{$^{198,199}$Os}\vspace{0.0cm}

The first refereed publication of the observation of $^{198}$Os and $^{199}$Os was the 2008 paper ``Single-particle behavior at N = 126: Isomeric decays in neutron-rich $^{204}$Pt'' by Steer et al. \cite{2008Ste01}. A 1 GeV/A $^{208}$Pb beam from the SIS-18 accelerator at GSI impinged on a $^9$Be target and the projectile fragments were selected and identified in-flight by the Fragment Separator FRS. The observation of the new neutron-rich osmium isotopes was not specifically mentioned but $^{198}$Os and $^{199}$Os events are clearly visible and identified in the particle identification plot in the first figure.

\section{Discovery of $^{165-202}$Ir}

Thirty-eight iridium isotopes from A = 165--202 have been discovered so far; these include 2 stable, 26 neutron-deficient and 10 neutron-rich isotopes. According to the HFB-14 model \cite{2007Gor01}, $^{258}$Ir should be the last odd-odd particle stable neutron-rich nucleus while the odd-even particle stable neutron-rich nuclei should continue through $^{261}$Ir. The proton dripline has already been crossed with the observation of the proton emitters $^{165}$Ir, $^{166}$Ir, and $^{167}$Ir, however, about six additional proton-rich iridium isotopes could still have half-lives longer than 10$^{-9}$~s \cite{2004Tho01}. Thus, about 63 isotopes have yet to be discovered corresponding to 62\% of all possible iridium isotopes.

In 2003, J.W. Arblaster published a review article entitled ``The Discoverers of the Iridium Isotopes'' \cite{2003Arb01}. Although he selected slightly different criteria for the discovery, our assignments agree in most of the cases. Since then only three additional isotopes ($^{200-202}$Ir) were discovered.

\begin{figure}
	\centering
	\includegraphics[scale=.7]{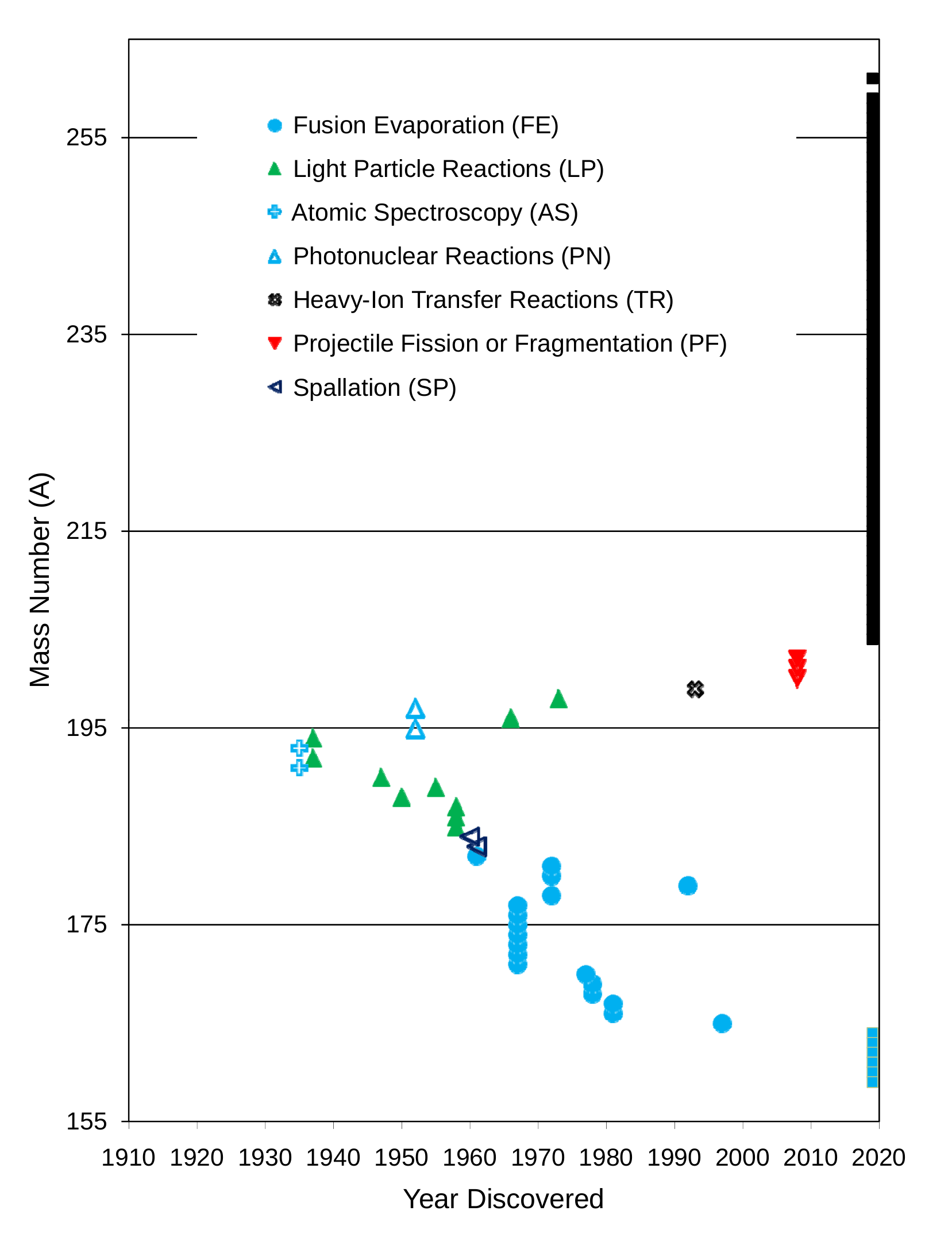}
	\caption{Iridium isotopes as a function of time when they were discovered. The different production methods are indicated. The solid black squares on the right hand side of the plot are isotopes predicted to be bound by the HFB-14 model. On the proton-rich side the light blue squares correspond to unbound isotopes predicted to have half-lives larger than $\sim 10^{-9}$~s.}
\label{f:year-ir}
\end{figure}

Figure \ref{f:year-ir} summarizes the year of first discovery for all iridium isotopes identified by the method of discovery. The range of isotopes predicted to exist is indicated on the right side of the figure. The radioactive iridium isotopes were produced using fusion evaporation reactions (FE), light-particle reactions (LP), photo-nuclear reactions (PN), heavy-ion induced transfer reactions (TR), spallation (SP), and projectile fission or fragmentation (PF). The stable isotopes were identified using atomic spectroscopy (AS). Light particles also include neutrons produced by accelerators. In the following, the discovery of each iridium isotope is discussed in detail.

\subsection*{$^{165}$Ir}\vspace{0.0cm}

$^{165}$Ir was discovered by Davids et al. as reported in the 1997 paper ``New proton radioactivities $^{165,166,167}$Ir and $^{171}$Au'' \cite{1997Dav01}. A 384~MeV $^{78}$Kr beam from the ATLAS accelerator was used to form $^{165}$Ir in the fusion-evaporation reaction $^{92}$Mo($^{78}$Kr,p4n). Charged particles were detected in double-sided silicon strip detectors at the end of the Fragment Mass Analyzer. ``One proton group is seen having an energy of 1707(7)~keV and a half-life of 0.29(6)~ms. Alpha events with energy 6715(7)~keV were also observed, correlated with two generations of known alphas from $^{161}$Re and $^{157}$Ta. The $^{165}$Ir alpha half-life was measured to be 0.39(16)~ms. Since one proton and one alpha group with similar half-lives were observed in the decay of $^{165}$Ir, both particles most likely come from the same state. The mean half-life is 0.30(6)~ms.'' This corresponds to the currently accepted half-life of an isomeric state.

\subsection*{$^{166,167}$Ir}\vspace{0.0cm}

The 1981 paper ``New neutron deficient isotopes in the range of elements Tm to Pt'' reported the discovery of $^{166}$Ir and $^{167}$Ir by Hofmann et al. at the linear accelerator UNILAC, GSI, Darmstadt in Germany \cite{1981Hof01}. $^{166}$Ir and $^{167}$Ir were produced in reactions bombarding neutron deficient targets between molybdenum and tin with a beam of $^{58}$Ni as well as bombarding targets between vanadium and nickel with a beam of $^{107}$Ag, with energies between 4.4 MeV/u and 5.9 MeV/u. Residues were separated using the velocity filter SHIP. ``The Ir isotopes with mass numbers 167 and 166 were produced in p2n and p3n reactions. Their $\alpha$ lines could be correlated to their Re daughter decays.'' The $\alpha$-decay energies of 6541(20)~keV and 6386(20)~keV for $^{166}$Ir and $^{167}$Ir, respectively are listed in a table. Only an upper limit of 5~ms could be determined for the half-lives.

\subsection*{$^{168,169}$Ir}\vspace{0.0cm}

``Copper ion induced reactions on $^{110-108-106}$Cd, $^{109-107}$Ag and $^{110}$Pd. New rhenium, osmium and iridium isotopes'' was published in 1978 by Cabot et al. announcing the discovery of $^{168}$Ir and $^{169}$Ir \cite{1978Cab01}. A 400~MeV $^{63}$Cu beam from the ALICE accelerator at Orsay, France, bombarded isotopically enriched cadmium targets to populate the iridium isotopes in the reactions $^{108}$Cd($^{63}$Cu,3n)$^{168}$Ir, $^{108}$Cd($^{63}$Cu,2n)$^{169}$Ir, and $^{110}$Cd($^{63}$Cu,4n)$^{169}$Ir. Alpha particles from fragments collected by a He-jet were detected to determine the decay energies and half-lives. ``The production curve for the 6.11~MeV $\alpha$-ray follows the (Cu,pn) excitation function and we attribute this activity to the decay of $^{169}$Ir formed by a (Cu,2n) reaction. Above E$^*$ = 55~MeV another new $\alpha$ peak is present in the $\alpha$ spectra at E$_{\alpha}$ = 6.22~MeV. Its yield curve corresponds to one more emitted particle and we attribute this activity to $^{168}$Ir since the (Cu,3n) is the more likely reaction.'' A half-life was only extracted for $^{169}$Ir. The measured value of 0.4(1)~s is close to the presently adopted value of 281(4)~ms for an isomeric state. Less than two months later Schrewe et al. reported independently a half-life of 0.4(2)~s for $^{169}$Ir \cite{1978Sch01}.

\subsection*{$^{170}$Ir}\vspace{0.0cm}

Cabot et al. published the first observation of $^{170}$Ir in the paper ``New osmium and iridium isotopes produced through $^{63}$Cu induced reactions'' in 1977 \cite{1977Cab01}. Self-supporting $^{110}$Cd targets were bombarded with a 380~MeV $^{63}$Cu beam from the Orsay ALICE accelerator. Reaction fragments were collected with a He-jet and $\alpha$-decay half-lives and decay energies were measured. ``New $\alpha$ active osmium and iridium isotopes $^{168}$Os, $^{167}$Os, $^{166}$Os, and $^{170}$Ir have been identified by cross bombardments and excitation functions measurements.'' The reported half-life of 1.1(2)~s is close to the currently adopted value of 811(18)~ms for a high-spin isomeric state.

\subsection*{$^{171-177}$Ir}\vspace{0.0cm}

In the paper entitled ``Alpha-active iridium isotopes'', Siivola described the discovery of $^{171}$Ir, $^{172}$Ir, $^{173}$Ir, $^{174}$Ir, $^{175}$Ir, $^{176}$Ir, and $^{177}$Ir  in 1967 \cite{1967Sii02}. The Berkeley Hilac accelerated $^{19}$F beams to 105$-$185~MeV which bombarded enriched targets of $^{162}$Er, $^{164}$Er, and $^{166}$Er. Alpha spectra were measured with a Au-Si surface barrier counter at the end of a continuously operating recoil collection apparatus. Alpha-decay energies and half-lives are listed in a table. The half-lives were: 1.0(3)~s ($^{171}$Ir), 1.7(5)~s ($^{172}$Ir), 3.0(10)~s ($^{173}$Ir), 4.0(10)~s ($^{174}$Ir), 4.5(10)~s ($^{175}$Ir), 8(1)~s ($^{176}$Ir), and 21(2)~s ($^{177}$Ir). The first four half-lives are consistent with the presently adopted values of isomeric states, and the half-lives of the three heaviest isotopes are at least within a factor of two of the adopted values for the corresponding ground states.

\subsection*{$^{178}$Ir}\vspace{0.0cm}

In 1972, Akhmadzhanov et al. reported the discovery of $^{178}$Ir in the paper ``The new isotopes $^{178}$Ir, $^{180}$Ir, $^{181}$Ir, decay scheme for $^{182}$Ir'' \cite{1972Akh01}. A 140~MeV $^{16}$O beam from the JINR U-300 accelerator bombarded thulium targets and $^{178}$Ir was formed in the fusion-evaporation reaction $^{169}$Tm($^{16}$O,7n). Gamma-rays singles, $\gamma-\gamma$- coincidence spectra and decay curves were measured. ``The $^{178}$Ir was obtained by irradiating metallic thulium with $^{16}$O ions having a maximum energy of 140 MeV for 0.5$-$1.0~min. Measurement of the $\gamma$-spectra was begun 5~sec after the end of irradiation. This isotope was identified on the basis of the $\gamma$-transitions 6$^+ \rightarrow  4^+ \rightarrow  2^+ \rightarrow  0^+$ between the levels of the ground-state band of the $^{178}$Os daughter nucleus, having energies of 363.1, 266.1, and 131.6~keV. The decay half-life of $^{178}$Ir determined from these $\gamma$-transitions is 0.5$\pm$0.3~min.'' This half-life is consistent with the currently accepted value 12(2)~s.

\subsection*{$^{179}$Ir}\vspace{0.0cm}

The observation of $^{179}$Ir was reported in the 1992 paper ``The decay of the isotopes $^{179}$Ir and $^{180}$Ir'' by Bosch-Wocke et al. \cite{1992Bos01}. Enriched $^{148}$Nd targets were irradiated with a 240~MeV $^{36}$Ar beam from the HMI Berlin VICKSI accelerator. X-ray and $\gamma$-ray time and energy spectra were recorded as singles and coincidences. ``The half-life analysis of $^{179}$Ir yielded t$_{1/2}$ = 79(1)~s from $\beta$-delayed $\gamma$-rays.'' This value corresponds to the presently accepted value. A previously reported half-life of 4(1)~min \cite{1971Nad02} was evidently incorrect.

\subsection*{$^{180,181}$Ir}\vspace{0.0cm}

In 1972, Akhmadzhanov et al. reported the discovery of $^{180}$Ir and $^{182}$Ir in the paper ``The new isotopes $^{178}$Ir, $^{180}$Ir, $^{181}$Ir, decay scheme for $^{182}$Ir'' \cite{1972Akh01}. $^{16}$O beams from the JINR U-300 accelerator bombarded thulium targets and $^{180}$Ir and $^{181}$Ir were formed in the fusion-evaporation reactions $^{169}$Tm($^{16}$O,5n) and $^{169}$Tm($^{16}$O,4n), respectively. Gamma-rays singles, $\gamma-\gamma$- coincidence spectra and decay curves were measured. ``$^{180}$Ir:... We obtained this iridium isotope by irradiating metallic thulium with 121~MeV $^{16}$O ions. The identification was based on the intensity decay of the $\gamma$-transitions with energies of 276.3 and 132.2~keV, which depopulate the levels of the known ground-state band of $^{180}$Os. The half-life found in this manner is 1.5$\pm$0.1~min... $^{181}$Ir: This isotope was identified on the basis of the genetic relation to the daughter $^{181}$Os (T$_{1/2}$ = 2.7~min and $^{181}$Os (105~min), whose decay has been studied thoroughly.'' The measured half-lives of 1.5(1)~min ($^{180}$Ir) and 5.0(3)~min ($^{181}$Ir) agree with the presently adopted values of 1.5(1)~min and 4.90(15)~min, respectively.  Previously reported half-lives 6.5(15)~min of 10(2)~min and for $^{180}$Ir and $^{181}$Ir, respectively, \cite{1971Nad02} were evidently incorrect.

\subsection*{$^{182}$Ir}\vspace{0.0cm}

Diamond et al. reported on the discovery of $^{182}$Ir in their 1961 publication ``The neutron deficient iridium isotopes Ir$^{182}$, Ir$^{183}$ and Ir $^{184}$'' \cite{1961Dia01}. The Berkeley heavy-ion linear accelerator Hilac was used to bombard metallic thulium targets with 160 MeV $^{16}$O ions. $^{182}$Ir was formed in the fusion-evaporation reaction $^{169}$Tm($^{16}$O,3n) and identified with end-window, flowing methane, proportional counters and two scintillation spectrometers following chemical separation. ``The decay curve of the iridium fraction from the Ir(O$^{16}$,xn) reaction is shown in [the figure]. The half life of the shortest component, which we assign to Ir$^{182}$, is 15$\pm$1~min.''  This half-life corresponds to the currently adopted value.

\subsection*{$^{183}$Ir}\vspace{0.0cm}

The discovery of $^{183}$Ir was reported in 1961 by Lavrukhina et al. in ``A new isotope Ir$^{183}$'' \cite{1961Lav01}. A 660~MeV proton beam from the Joint Institute for Nuclear Studies synchrocyclotron bombarded a metallic gold target. Decay curves were measured with an end-window counter following chemical separation. ``The variation of the Os$^{183}$ activity as a function of extraction time gives a half-life of Ir$^{183}$ equal to (1$\pm$0.1)~hr.'' This half-life agrees with the presently adopted value of 58(5)~min. Diamond et al. reported independently a half-life of 55(7)~min 5 months later \cite{1961Dia01}. A previously measured half-life of ``not less than 5 hours'' \cite{1960Sur01} was evidently incorrect.

\subsection*{$^{184}$Ir}\vspace{0.0cm}

Baranov et al. reported the first identification of $^{184}$Ir in the 1961 article ``New iridium and platinum isotopes: Ir$^{184}$ and Pt$^{187}$'' \cite{1960Bar01}. $^{184}$Ir was produced by bombarding a gold target with 660 MeV protons from the Dubna JINR synchrocyclotron and identified with a Danysz type $\beta$-spectrometer following chemical separation. ``We discovered a new iridium isotope - Ir$^{184}$ - with a period of 3.1$\pm$0.3~hrs.'' This half-life agrees with the currently accepted value of 3.09(3)~h. A few months later Diamond et al. independently reported a half-life of 3.2(2)~h \cite{1961Dia01}.

\subsection*{$^{185-187}$Ir}\vspace{0.0cm}

In the 1958 paper ``Neutron-deficient iridium isotopes'', Diamond and Hollander reported the discovery of $^{185}$Ir, $^{186}$Ir and $^{187}$Ir  \cite{1958Dia01}. Natural rhenium targets were bombarded with $\alpha$ particles between 25 and 45~MeV from the Berkeley Crocker 60-inch cyclotron. Following chemical separation, decay curves were measured with proportional counter and $\gamma$-rays spectra were recorded with a sodium iodide crystal.  ``With alpha particles of 45~MeV initial energy Ir$^{185}$ is produced from rhenium by the reaction Re$^{185}$($\alpha$,4n)Ir$^{185}$; there are simultaneously produced the heavier isotopes of iridium from lower order reactions, The Ir$^{185}$ radiations can be distinguished, and the mass assignment made, by repeating the bombardment with alpha particles of initial energy lower than the threshold for the ($\alpha$,4n) reaction, i.e., 33$-$34~MeV, and observing which conversion lines and photons are eliminated at the lower energy irradiations... In this work we have established the transitions that belong to mass 186 by an observation of the particular electron lines and photons that disappear from the complex spectrum when the alpha particle bombarding energy is lowered from 33-34~MeV (threshold for production of Ir$^{185}$) to 25$-$27~MeV (threshold for production of Ir$^{186}$); such transitions are assigned to Ir$^{186}$... Iridiun-187: In irradiations of rhenium foils at an initial alpha particle energy of 25 to 27 MeV, i.e. below the threshold for the production of Ir$^{185}$ and Ir$^{186}$, several short-lived conversion electron lines and gamma rays were observed in the separated iridium fraction. By comparing the intensities of the more prominent of these electron lines in a series of timed exposures in the electron spectrographs with lines of known intensities in a group of standard plates, we obtained a value of 13$\pm$3 hours for the half-life of this new activity.'' The measured half-lives of 15(3)~h ($^{185}$Ir), 16(3)~h ($^{186}$Ir) and 13(3)~h ($^{187}$Ir) agree with the presently accepted values of 14.4(1)~h, 16.64(3)~h, and 10.5(3)~h, respectively. Earlier, Smith and Hollander had assigned a half-life of 14~h incorrectly to $^{187}$Ir \cite{1955Smi01}. Smith and Hollander also questioned the previous assignment of an 11.8~h half-life to $^{187}$Ir by Chu \cite{1950Chu01} arguing that it corresponded most likely to $^{186}$Ir.

\subsection*{$^{188}$Ir}\vspace{0.0cm}

The discovery of $^{188}$Ir was reported in the 1950 paper ``New radioactive isotopes of iridium'' by Chu \cite{1950Chu01}. Natural and enriched rhenium targets were bombarded with 38~MeV helium ions from the Berkeley Crocker 60-inch cyclotron. Decay curves and absorption spectra were measured with ``end on'' type argon-ethanol filled Geiger counters. ``41.5-Hr Ir$^{188}$: This isotope was produced both from $\alpha$-bombardment on rhenium and from deuteron bombardment on osmium. A study of its yield from natural rhenium targets justifies postulating its formation by an ($\alpha$,n) reaction on Re$^{185}$ and an ($\alpha$,3n) reaction on Re$^{187}$.'' The measured half-life of 41.5~h agrees with the presently accepted values of 41.5(5)~h.

\subsection*{$^{189}$Ir}\vspace{0.0cm}

Smith and Hollander first observed $^{189}$Ir in 1955 as reported in ``Radiochemical study of neutron-deficient chains in the noble metal region'' \cite{1955Smi01}. The experiments were performed with protons from the Berkeley 184-inch and 60-inch cyclotrons. Decay curves and $\gamma$-ray spectra were measured with a Geiger counter and NaI(Tl) scintillation spectrometer, respectively. ``Bombardments of iridium with 32-Mev protons produce in the platinum fraction active isotopes of masses 193, 191, 189, and 188. If one allows this fraction to decay for several weeks and then removes iridium from it, good samples of Ir$^{189}$ and Ir$^{188}$ are obtained, because Pt$^{191}$ and Pt$^{193m}$ have no active iridium daughters. The 11-day Ir$^{189}$ can be distinguished from 41-hour Ir$^{188}$ by virtue of their very different half-lives.'' This half-life agrees with the currently adopted value of 13.2(1)~h. Previously, Chu had assigned a 12.6~d half-life incorrectly as an isomer of $^{190}$Ir \cite{1950Chu01}.

\subsection*{$^{190}$Ir}\vspace{0.0cm}

In the paper ``Radioactive isotopes of Re, Os, and Ir'', Goodman and Pool described their discovery of $^{190}$Ir in 1947 \cite{1947Goo01}. $^{190}$Ir was produced in (d,n) reactions on osmium and (n,2n) reactions on iridium. Decay curves were measured following chemical separation. ``Since the 10.7-day period has been made by fast neutrons on iridium and by Os(d,n) and has not been found with slow neutrons or with deuterons on iridium, the activity is therefore ascribed to Ir$^{190}$.'' This half life agrees with the accepted value of 11.78(10)~d.

\subsection*{$^{191}$Ir}\vspace{0.0cm}

In 1935, Venkatesachar and Sibaiya discovered $^{191}$Ir as reported in the paper ``Iridium isotopes and their nuclear spin'' \cite{1935Ven01}. Arc lines of iridium radiated from a hollow cathode were analyzed at Central College in Bangalore. The hyperfine structure pattern was obtained with a Hilger quartz Lummer plate. ``The observed structure is accounted for uniquely by assuming two isotopes of masses 191 and 193 with nuclear spins 1/2 and 3/2, respectively... Iridium is one of the few elements the isotopic constitution of which has not so far been revealed by the mass-spectrograph.''

\subsection*{$^{192}$Ir}\vspace{0.0cm}

McMillan et al. identified $^{192}$Ir for the first time in 1937 in ``Neutron-induced radioactivity of the noble metals'' \cite{1937McM01}. Following the irradiation of a iridium target with slow and fast neutrons produced with a deuteron beam on lithium, activities of 2 months, 19 hr, and 1.5 min were observed. ``The 1.5-min. period is present with a saturation intensity of 0.2 div./sec., and the 19-hr. period is buried in the midst of a continuously curving logarithmic plot, so that we cannot be sure of its present. It is certainly less intense relative to the 2-month period than with slow neutron activation, just as is the 1.5-min. period, so that we can provisionally assign the 2-month period to Ir$^{192}$ and the other two to Ir$^{194}$.'' The half-life of 2 months for $^{192}$Ir is in reasonable agreement with the accepted value of 73.827(13)~d.  A 2~h half-life had been previously reported by Amaldi and Fermi without a specific mass assignment \cite{1936Ama01}.

\subsection*{$^{193}$Ir}\vspace{0.0cm}

In 1935, Venkatesachar and Sibaiya discovered $^{193}$Ir as reported in the paper ``Iridium isotopes and their nuclear spin'' \cite{1935Ven01}. Arc lines of iridium radiated from a hollow cathode were analyzed at Central College in Bangalore. The hyperfine structure pattern was obtained with a Hilger quartz Lummer plate. ``The observed structure is accounted for uniquely by assuming two isotopes of masses 191 and 193 with nuclear spins 1/2 and 3/2, respectively... Iridium is one of the few elements the isotopic constitution of which has not so far been revealed by the mass-spectrograph.''

\subsection*{$^{194}$Ir}\vspace{0.0cm}

McMillan et al. identified $^{194}$Ir for the first time in 1937 in ``Neutron-induced radioactivity of the noble metals'' \cite{1937McM01}. Following the irradiation of a iridium target with slow and fast neutrons produced with a deuteron beam on lithium, activities of 2 months, 19 hr, and 1.5 min were observed. ``The 1.5-min. period is present with a saturation intensity of 0.2 div./sec., and the 19-hr. period is buried in the midst of a continuously curving logarithmic plot, so that we cannot be sure of its presence. It is certainly less intense relative to the 2-month period than with slow neutron activation, just as is the 1.5-min. period, so that we can provisionally assign the 2-month period to Ir$^{192}$ and the other two to Ir$^{194}$.'' The half-life of 19~h for $^{194}$Ir is in agreement with the accepted value of 19.28(3)~h. A 20~h half-life had been previously reported by Fermi et al. without a specific mass assignment \cite{1934Fer01}.

\subsection*{$^{195}$Ir}\vspace{0.0cm}

The discovery of $^{195}$Ir was described in the 1952 paper ``Radioactivities of platinum and iridium from photonuclear reactions in platinum'' by Christian et al. \cite{1952Chr01}. Platinum samples were irradiated with X-rays from the Iowa State 70-MeV synchrotron. Decay curves were measured with mica end-window G-M tubes following chemical separation. ``The iridium fraction included Ir$^{192}$ and Ir$^{194}$ and two new isotopes: a 140-min, 1-Mev $\beta^-$ emitter, probably Ir$^{195}$, and a 7-min activity, probably Ir$^{197}$.'' The half-life assigned to $^{195}$Ir agrees with the presently adopted value of 2.5(2)~h. Previously, Butement had assigned a 66-min half-life to either $^{195}$Ir or $^{197}$Ir \cite{1951But01}.

\subsection*{$^{196}$Ir}\vspace{0.0cm}

In 1966, Vonach et al. discovered $^{196}$Ir as reported in the paper ``Untersuchung der bisher Ir$^{198}$ zugeschriebenen 50 sec-Aktivit\"at und Neuzuordnung zum Zerfall von Ir$^{196}$'' \cite{1966Von01}. Natural platinum and $^{196}$Pt enriched targets were irradiated with 14 MeV neutrons. $^{196}$Ir was identified by measuring $\beta$-, $\gamma$- and $\gamma-\gamma$-coincidence spectra. ``In this way the 50 sec activity assigned so far to Ir$^{198}$ could be identified as Ir$^{196}$.'' The measured half-life of 50(2)~s agrees with the currently adopted value of 52(1)~s. The previously incorrect assignment to $^{198}$Ir mentioned in the quote was made by Butement and Poe who had assigned a 9.7~d half-life to $^{196}$Ir \cite{1954But01}. Also previously, Bishop had assigned a 2~h half-life to $^{196}$Ir \cite{1965Bis01}, however, Jansen and Pauw suggested that Bishop had observed a mixture of $^{196}$Ir and $^{195}$Ir \cite{1967Jan01}.

\subsection*{$^{197}$Ir}\vspace{0.0cm}

The discovery of $^{197}$Ir was described in the 1952 paper ``Radioactivities of platinum and iridium from photonuclear reactions in platinum'' by Christian et al. \cite{1952Chr01}. Platinum samples were irradiated with X-rays from the Iowa State 70-MeV synchrotron. Decay curves were measured with mica end-window G-M tubes following chemical separation. ``The iridium fraction included Ir$^{192}$ and Ir$^{194}$ and two new isotopes: a 140-min, 1-Mev $\beta^-$ emitter, probably Ir$^{195}$, and a 7-min activity, probably Ir$^{197}$.'' The half-life assigned to $^{197}$Ir agrees with the presently adopted value of 5.8(5)~min. Previously, Butement had assigned a 66-min half-life to either $^{195}$Ir or $^{197}$Ir \cite{1951But01}.

\subsection*{$^{198}$Ir}\vspace{0.0cm}

Szalay and Uray reported the discovery of $^{198}$Ir in the 1973 paper ``Evidence for the existence of $^{198}$Ir'' \cite{1973Sza01}. Natural platinum and $^{198}$Pt enriched targets were irradiated with 14 MeV neutrons. $^{198}$Ir was identified by comparing decay curves and $\gamma$-ray spectra from the two targets.  ``At about 10 sec neutron activation period a peak with half-life of 8$\pm$3 sec appeared in the $\gamma$-ray spectrum of the activated enriched Pt target at 407.76$\pm$0.22 keV... The measured data do not enable to decide, whether this half-life belongs to the ground state or to a metastable state of $^{198}$Ir, or it is a mixture of two comparable ones. Additional information can be drawn nevertheless from the systematics of neighbouring nuclei, which indicate that this half-life belongs to the ground state of $^{198}$Ir.'' This half-life agrees with the presently adopted value of 8(1)~s.  Previously, Butement and Poe had incorrectly assigned the 50~s half-life of $^{196}$Ir to $^{198}$Ir \cite{1954But01}.

\subsection*{$^{199}$Ir}\vspace{0.0cm}

$^{199}$Ir was discovered in 1995 by Zhao et al. as reported in their paper ``Production of $^{199}$Ir via exotic nucleon transfer reaction'' \cite{1993Zha01}. A 140~MeV $^{18}$O beam was used to bombard enriched $^{198}$Pt targets and $^{199}$Ir was produced in the heavy-ion transfer reaction $^{198}$Pt($^{18}$O,$^{17}$F). Reaction fragments were measured and with a high resolution QMG/2 magnetic spectrometer at Daresbury, England. ``The experimental results have shown that although the yield is very low, there is good evidence for a distinct upper limit to the spectrum. Taking this point as the position of the ground state, gives a Q-value for the reaction of $-$8.241$\pm$0.034~MeV, and a mass excess for $^{199}$Ir of $-$24.424$\pm$0.034~MeV.''

\subsection*{$^{200-202}$Ir}\vspace{0.0cm}

The first refereed publication of the observation of $^{200}$Ir, $^{201}$Ir, and $^{202}$Ir was the 2008 paper ``Single-particle behavior at N = 126: isomeric decays in neutron-rich $^{204}$Pt'' by Steer et al. \cite{2008Ste01}. A 1 GeV/A $^{208}$Pb beam from the SIS-18 accelerator at GSI impinged on a $^9$Be target and the projectile fragments were selected and identified in-flight by the Fragment Separator FRS. The observation of the new neutron-rich iridium isotopes was not specifically mentioned but $^{200}$Ir, $^{201}$Ir, and $^{202}$Ir events are clearly visible and identified in the particle identification plot in the first figure.

\section{Summary}
The discoveries of the known tantalum, rhenium, osmium, and iridium isotopes have been compiled and the methods of their production discussed. The identification of these isotopes was relatively easy with only a few isotopes initially incorrectly identified.

The first reports of the half-lives of $^{155}$Ta, $^{167}$Ta, $^{172}$ta, and $^{180}$Ta were incorrect and the half-lives of $^{180}$Ta and $^{182}$Ta were first reported without a mass assignment.

The rhenium isotopes $^{167}$Re, $^{178}$Re, $^{186}$Re, and $^{188}$Re were at first incorrectly identified, the half-life of $^{183}$Re initially not correct, and the half-lives of $^{173}$Re, $^{186}$Re, and $^{188}$Re could not be assigned to a specific isotope.

$^{180}$Os had initially been assigned to $^{181}$Os, no mass assignment was made for the half-life of $^{193}$Os, and the first report of the $^{195}$Os half-life was at first incorrect.

The original half-lives reported for $^{179}$Ir, $^{180}$Ir, $^{181}$Ir, $^{183}$Ir, and $^{196}$Ir were incorrect. The first mass assignments of $^{187}$Ir and $^{189}$Ir were not accurate and the half-lives of $^{192}$Ir, $^{194}$Ir, and $^{195}$Ir were reported without mass assignments.

The discovery of the osmium \cite{2004Arb01} and iridium \cite{2003Arb01} isotopes was discussed several years ago by Arblaster. For the osmium isotopes the present assignments agree with the assignments by Arblaster with the exception of $^{178}$Os. While Arblaster credits the first half-life measurement by Belyaev et al. in 1968 \cite{1968Bel01}, we recognize the 1967 measurement of the ground-state rotational levels by Burde et al. \cite{1967Bur01}.

Also, most of the assignments for the iridium isotopes agree with assignments by Arblaster \cite{2003Arb01}. The exceptions are: (1) $^{164}$Ir, which we did not include in the current compilation because it so far has only been reported in conference proceedings \cite{2001Ket02,2002Mah01}; (2) for $^{179}$Ir the observation by Nadzhakov et al. \cite{1971Nad02} was accepted despite the wrong half-life; (3) $^{186}$Ir was assigned to a private communication by Scharff-Goldhaber mentioned in reference \cite{1958Dia01}; (4) the first observation of the two stable isotopes $^{191}$Ir and $^{193}$Ir by Venkatesachar and Sibaiya \cite{1935Ven01} was not accepted because it apparently contradicted the atomic weight known at the time; (5) an unpublished report \cite{1972Sch02} was credited for the observation of $^{198}$Ir.

\ack

This work was supported by the National Science Foundation under grant No. PHY06-06007 (NSCL).

\bibliography{../isotope-discovery-references}

\newpage

\newpage

\TableExplanation

\bigskip
\renewcommand{\arraystretch}{1.0}

\section{Table 1.\label{tbl1te} Discovery of tantalum, rhenium, osmium, and iridium isotopes }
\begin{tabular*}{0.95\textwidth}{@{}@{\extracolsep{\fill}}lp{5.5in}@{}}
\multicolumn{2}{p{0.95\textwidth}}{ }\\

Isotope & Tantalum, rhenium, osmium, and iridium isotope \\
Author & First author of refereed publication \\
Journal & Journal of publication \\
Ref. & Reference \\
Method & Production method used in the discovery: \\

  & FE: fusion evaporation \\
  & LP: light-particle reactions (including neutrons) \\
  & MS: mass spectroscopy \\
  & AS: atomic spectroscopy \\
  & NC: neutron capture reactions \\
  & PN: photo-nuclear reactions \\
  & TR: heavy-ion induced transfer reactions \\
  & SP: spallation \\
  & PF: projectile fission or fragmentation \\

Laboratory & Laboratory where the experiment was performed\\
Country & Country of laboratory\\
Year & Year of discovery \\
\end{tabular*}
\label{tableI}

\datatables 



\setlength{\LTleft}{0pt}
\setlength{\LTright}{0pt}


\setlength{\tabcolsep}{0.5\tabcolsep}

\renewcommand{\arraystretch}{1.0}

\footnotesize 

\begin{longtable}{@{\extracolsep\fill}llllllll@{}}
\caption{Discovery of Tantalum, Rhenium, Osmium, and Iridium Isotopes. See page\ \pageref{tbl1te} for Explanation of Tables}
Isotope & Author & Journal & Ref. & Method & Laboratory & Country & Year\\
\hline\\
\endfirsthead\\
\caption[]{(continued)}
Isotope & Author & Journal & Ref. & Method & Laboratory & Country & Year\\
\hline\\
\endhead
$^{155}$Ta& R.D. Page & Phys. Rev. C &\cite{2007Pag01}& FE & Jyv\"askyl\"a & Finland &2007 \\
$^{156}$Ta & R.D. Page & Phys. Rev. Lett. &\cite{1992Pag01}& FE & Daresbury & UK &1992 \\
$^{157}$Ta & S. Hofmann & Z. Phys. A &\cite{1979Hof01}& FE & Darmstadt & Germany &1979 \\
$^{158}$Ta & S. Hofmann & Z. Phys. A &\cite{1979Hof01}& FE & Darmstadt & Germany &1979 \\
$^{159}$Ta & S. Hofmann & Z. Phys. A &\cite{1979Hof01}& FE & Darmstadt & Germany &1979 \\
$^{160}$Ta & S. Hofmann & Z. Phys. A &\cite{1979Hof01}& FE & Darmstadt & Germany &1979 \\
$^{161}$Ta & S. Hofmann & Z. Phys. A &\cite{1979Hof01}& FE & Darmstadt & Germany &1979 \\
$^{162}$Ta & C.F. Liang & Z. Phys. A &\cite{1985Lia01}& SP & Orsay & France &1985 \\
$^{163}$Ta & C.F. Liang & Z. Phys. A &\cite{1985Lia01}& SP & Orsay & France &1985 \\
$^{164}$Ta & B. Eichler & Radiochem. Radioanal. Lett. &\cite{1982Eic01}& FE & Dubna & Russia &1982 \\
$^{165}$Ta & H. Bruchertseifer & Radiochem. Radioanal. Lett. &\cite{1982Bru01}& FE & Dubna & Russia &1982 \\
$^{166}$Ta & R.E. Leber & J. Inorg. Nucl. Chem. &\cite{1977Leb01}& FE & Yale & USA &1977 \\
$^{167}$Ta & C.F. Liang & Z. Phys. A &\cite{1982Lia01}& SP & Orsay & France &1982 \\
$^{168}$Ta & R. Arlt & Bull. Acad. Sci. USSR &\cite{1969Arl03}& SP & Dubna & Russia &1969 \\
$^{169}$Ta & R. Arlt & Bull. Acad. Sci. USSR &\cite{1969Arl03}& SP & Dubna & Russia &1969 \\
$^{170}$Ta & R. Arlt & Bull. Acad. Sci. USSR &\cite{1969Arl03}& SP & Dubna & Russia &1969 \\
$^{171}$Ta & R. Arlt & Bull. Acad. Sci. USSR &\cite{1969Arl03}& SP & Dubna & Russia &1969 \\
$^{172}$Ta & H. Abou-Leila & Phys. Rev. C &\cite{1964Abo01}& LP & Orsay & France &1964 \\
$^{173}$Ta & K.T. Faler & Phys. Rev. &\cite{1960Fal01}& FE & Berkeley & USA &1960 \\
$^{174}$Ta & K.T. Faler & Phys. Rev. &\cite{1960Fal01}& FE & Berkeley & USA &1960 \\
$^{175}$Ta & K.T. Faler & Phys. Rev. &\cite{1960Fal01}& FE & Berkeley & USA &1960 \\
$^{176}$Ta & G. Wilkinson & Phys. Rev. &\cite{1948Wil02}& LP & Berkeley & USA &1948 \\
$^{177}$Ta & G. Wilkinson & Phys. Rev. &\cite{1948Wil02}& LP & Berkeley & USA &1948 \\
$^{178}$Ta & G. Wilkinson & Phys. Rev. &\cite{1950Wil01}& LP & Berkeley & USA &1950 \\
$^{179}$Ta & G. Wilkinson & Phys. Rev. &\cite{1950Wil01}& LP & Berkeley & USA &1950 \\
$^{180}$Ta & O. Oldenberg & Phys. Rev. &\cite{1938Old01}& NC & Berkeley & USA &1938 \\
$^{181}$Ta & F.W. Aston & Nature &\cite{1932Ast01}& MS & Cambridge & UK &1932 \\
$^{182}$Ta & O. Oldenberg & Phys. Rev. &\cite{1938Old01}& NC & Berkeley & USA &1938 \\
$^{183}$Ta & F.D.S. Butement & Nature &\cite{1950But01}& PN & Harwell & UK &1950 \\
$^{184}$Ta & F.D.S. Butement & Phil. Mag. &\cite{1955But02}& LP & Harwell & UK &1955 \\
$^{185}$Ta & F.D.S. Butement & Nature &\cite{1950But01}& PN & Harwell & UK &1950 \\
$^{186}$Ta & A.J. Poe & Phil. Mag. &\cite{1955Poe01}& LP & Harwell & UK &1955 \\
$^{187}$Ta & J. Benlliure & Nucl. Phys. A &\cite{1999Ben01}& PF & Darmstadt & Germany &1999 \\
$^{188}$Ta & J. Benlliure & Nucl. Phys. A &\cite{1999Ben01}& PF & Darmstadt & Germany &1999 \\
$^{189}$Ta & J. Benlliure & Nucl. Phys. A &\cite{1999Ben01}& PF & Darmstadt & Germany &1999 \\
$^{190}$Ta & N. Alkhomashi & Phys. Rev. C &\cite{2009Alk01}& PF & Darmstadt & Germany &2009 \\
$^{191}$Ta & N. Alkhomashi & Phys. Rev. C &\cite{2009Alk01}& PF & Darmstadt & Germany &2009 \\
$^{192}$Ta & N. Alkhomashi & Phys. Rev. C &\cite{2009Alk01}& PF & Darmstadt & Germany &2009 \\
 & & & & & & & \\
 & & & & & & & \\
$^{159}$Re& D.T. Joss & Phys. Lett. B &\cite{2006Jos01}& FE & Jyv\"askyl\"a & Finland &2006 \\
$^{160}$Re& R.D. Page & Phys. Rev. Lett. &\cite{1992Pag01}& FE & Daresbury & UK &1992 \\
$^{161}$Re& S. Hofmann & Z. Phys. A &\cite{1979Hof01}& FE & Darmstadt & Germany &1979 \\
$^{162}$Re& S. Hofmann & Z. Phys. A &\cite{1979Hof01}& FE & Darmstadt & Germany &1979 \\
$^{163}$Re& S. Hofmann & Z. Phys. A &\cite{1979Hof01}& FE & Darmstadt & Germany &1979 \\
$^{164}$Re& S. Hofmann & Z. Phys. A &\cite{1979Hof01}& FE & Darmstadt & Germany &1979 \\
$^{165}$Re& S. Hofmann & Z. Phys. A &\cite{1981Hof01}& FE & Darmstadt & Germany &1981 \\
$^{166}$Re& U.J. Schrewe & Z. Phys. A &\cite{1978Sch01}& FE & Darmstadt & Germany &1978 \\
$^{167}$Re & F. Meissner & Z. Phys. A &\cite{1992Mei01}& FE & Berlin & Germany &1992 \\
$^{168}$Re & F. Meissner & Z. Phys. A &\cite{1992Mei01}& FE & Berlin & Germany &1992 \\
$^{169}$Re & C. Cabot & Z. Phys. A &\cite{1978Cab01}& FE & Orsay & France &1978 \\
$^{170}$Re & E.E. Berlovich & Bull. Acad. Sci. USSR &\cite{1974Ber01}& SP & Leningrad & Russia &1974 \\
$^{171}$Re & E. Runte & Z. Phys. A &\cite{1987Run02}& FE & Berlin & Germany &1987 \\
$^{172}$Re & E.E. Berlovich & Bull. Acad. Sci. USSR &\cite{1972Ber01}& SP & Leningrad & Russia &1972 \\
$^{173}$Re & A. Szymanski & Radiochim. Acta &\cite{1986Szy01}& FE & Manchester & UK &1986 \\
$^{174}$Re & E.E. Berlovich & Bull. Acad. Sci. USSR &\cite{1972Ber01}& SP & Leningrad & Russia &1972 \\
$^{175}$Re & E. Nadjakov & Compt. Rend. Acad. Bulgare Sci. &\cite{1967Nad01}& FE & Dubna & Russia &1967 \\
$^{176}$Re & E. Nadjakov & Compt. Rend. Acad. Bulgare Sci. &\cite{1967Nad01}& FE & Dubna & Russia &1967 \\
$^{177}$Re & B.C. Haldar & Phys. Rev. &\cite{1957Hal01}& LP & Rochester & USA &1957 \\
$^{178}$Re & B.C. Haldar & Phys. Rev. &\cite{1957Hal01}& LP & Rochester & USA &1957 \\
$^{179}$Re & B. Harmatz & Phys. Rev. &\cite{1960Har01}& LP & Oak Ridge & USA &1960 \\
$^{180}$Re & V. Kistiakowsky & Phys. Rev. &\cite{1955Kis01}& LP & Berkeley & USA &1955 \\
$^{181}$Re & C.J. Gallagher & Phys. Rev. &\cite{1957Gal01}& LP & Berkeley & USA &1957 \\
$^{182}$Re & G. Wilkinson & Phys. Rev. &\cite{1950Wil02}& LP & Berkeley & USA &1950 \\
$^{183}$Re & B.J. Stover & Phys. Rev. &\cite{1950Sto01}& LP & Berkeley & USA &1950 \\
$^{184}$Re & K, Fajans & Phys. Rev. &\cite{1940Faj01}& LP & Michigan & USA &1940 \\
$^{185}$Re & F.W. Aston & Nature &\cite{1931Ast04}& MS & Cambridge & UK &1931 \\
$^{186}$Re & K. Sinma & Phys. Rev. &\cite{1939Sin01}& NC & Tokyo & Japan &1939 \\
$^{187}$Re & F.W. Aston & Nature &\cite{1931Ast04}& MS & Cambridge & UK &1931 \\
$^{188}$Re & K. Sinma & Phys. Rev. &\cite{1939Sin01}& NC & Tokyo & Japan &1939 \\
$^{189}$Re & B. Crasemann & Phys. Rev. &\cite{1963Cra01}& LP & Brookhaven & USA &1963 \\
$^{190}$Re & A.H.W. Aten Jr.& Physica &\cite{1955Ate03}& LP & Amsterdam & Netherlands &1955 \\
$^{191}$Re & B. Crasemann & Phys. Rev. &\cite{1963Cra01}& LP & Brookhaven & USA &1963 \\
$^{192}$Re & J. Blachot & Compt. Rend. Acad. Sci. &\cite{1965Bla01}& LP & Grenoble & France &1965 \\
$^{193}$Re & J. Benlliure & Nucl. Phys. A &\cite{1999Ben01}& PF & Darmstadt & Germany &1999 \\
$^{194}$Re & J. Benlliure & Nucl. Phys. A &\cite{1999Ben01}& PF & Darmstadt & Germany &1999 \\
$^{195}$Re & S.T. Steer & Phys. Rev. &\cite{2008Ste01}& PF & Darmstadt & Germany &2008 \\
$^{196}$Re & S.T. Steer & Phys. Rev. &\cite{2008Ste01}& PF & Darmstadt & Germany &2008 \\
 & & & & & & & \\
 & & & & & & & \\
$^{161}$Os& L. Bianco & Phys. Lett. B &\cite{2010Bia01}& FE & Jyv\"askyl\"a & Finland &2010 \\
$^{162}$Os& S. Hofmann & Z. Phys. A &\cite{1989Hof01}& FE & Darmstadt & Germany &1989 \\
$^{163}$Os& S. Hofmann & Z. Phys. A &\cite{1981Hof01}& FE & Darmstadt & Germany &1981 \\
$^{164}$Os& S. Hofmann & Z. Phys. A &\cite{1981Hof01}& FE & Darmstadt & Germany &1981 \\
$^{165}$Os& C. Cabot & Z. Phys. A &\cite{1978Cab01}& FE & Orsay & France &1978 \\
$^{166}$Os & C. Cabot & Z. Phys. A &\cite{1977Cab01}& FE & Orsay & France &1977 \\
$^{167}$Os & C. Cabot & Z. Phys. A &\cite{1977Cab01}& FE & Orsay & France &1977 \\
$^{168}$Os & C. Cabot & Z. Phys. A &\cite{1977Cab01}& FE & Orsay & France &1977 \\
$^{169}$Os & K.S. Toth & Phys. Rev. C &\cite{1972Tot01}& FE & Oak Ridge & USA &1972 \\
$^{170}$Os & K.S. Toth & Phys. Rev. C &\cite{1972Tot02}& FE & Oak Ridge & USA &1972 \\
$^{171}$Os & K.S. Toth & Phys. Rev. C &\cite{1972Tot02}& FE & Oak Ridge & USA &1972 \\
$^{172}$Os & J. Borggreen & Nucl. Phys. A &\cite{1971Bor01}& FE & Berkeley & USA &1971 \\
$^{173}$Os & J. Borggreen & Nucl. Phys. A &\cite{1971Bor01}& FE & Berkeley & USA &1971 \\
$^{174}$Os & J. Borggreen & Nucl. Phys. A &\cite{1971Bor01}& FE & Berkeley & USA &1971 \\
$^{175}$Os & E.E. Berlovich & Bull. Acad. Sci. USSR &\cite{1972Ber01}& SP & Leningrad & Russia &1972 \\
$^{176}$Os & R. Arlt & Bull. Acad. Sci. USSR &\cite{1970Arl01}& SP & Dubna & Russia &1970 \\
$^{177}$Os & R. Arlt & Bull. Acad. Sci. USSR &\cite{1970Arl01}& SP & Dubna & Russia &1970 \\
$^{178}$Os & J. Burde & Nucl. Phys. A &\cite{1967Bur01}& FE & Berkeley & USA &1967 \\
$^{179}$Os & B.N. Belyaev & Bull. Acad. Sci. USSR &\cite{1968Bel01}& FE & Dubna & Russia &1968 \\
$^{180}$Os & B.N. Belyaev & Sov. J. Nucl. Phys. &\cite{1967Bel01}& FE & Dubna & Russia &1967 \\
$^{181}$Os & K.J. Hofstetter & Phys. Rev. &\cite{1966Hof01}& LP & Argonne/ Oak Ridge& USA &1966 \\
$^{182}$Os & B.J. Stover & Phys. Rev. &\cite{1950Sto01}& LP & Berkeley & USA &1950 \\
$^{183}$Os & B.J. Stover & Phys. Rev. &\cite{1950Sto01}& LP & Berkeley & USA &1950 \\
$^{184}$Os & A.O. Nier & Phys. Rev. &\cite{1937Nie01}& MS & Harvard & USA &1937 \\
$^{185}$Os & L.J. Goodman & Phys. Rev. &\cite{1947Goo01}& LP & Ohio State & USA &1947 \\
$^{186}$Os & F.W. Aston & Nature &\cite{1931Ast03}& MS & Cambridge & UK &1931 \\
$^{187}$Os & F.W. Aston & Nature &\cite{1931Ast03}& MS & Cambridge & UK &1931 \\
$^{188}$Os & F.W. Aston & Nature &\cite{1931Ast03}& MS & Cambridge & UK &1931 \\
$^{189}$Os & F.W. Aston & Nature &\cite{1931Ast03}& MS & Cambridge & UK &1931 \\
$^{190}$Os & F.W. Aston & Nature &\cite{1931Ast03}& MS & Cambridge & UK &1931 \\
$^{191}$Os & E. Zingg& Helv. Phys. Acta &\cite{1940Zin01}& NC & Zurich & Switzerland &1940 \\
$^{192}$Os & F.W. Aston & Nature &\cite{1931Ast03}& MS & Cambridge & UK &1931 \\
$^{193}$Os & E. Zingg& Helv. Phys. Acta &\cite{1940Zin01}& NC & Zurich & Switzerland &1940 \\
$^{194}$Os & M. Lindner & Phys. Rev. &\cite{1951Lin02}& NC & Washington State & USA &1951 \\
$^{195}$Os & J.J. Valiente-Dobon & Phys. Rev. C &\cite{2004Val01}& PF & Darmstadt & Germany &2004 \\
$^{196}$Os & P.E. Haustein & Phys. Rev. C &\cite{1977Hau01}& LP & Brookhaven & USA &1977 \\
$^{197}$Os & Y. Xu & J. Radioanal. Nucl. Chem. &\cite{2003Xu01}& LP & Lanzhou & China &2003 \\
$^{198}$Os & S.T. Steer & Phys. Rev. C &\cite{2008Ste01}& PF & Darmstadt & Germany &2008 \\
$^{199}$Os & S.T. Steer & Phys. Rev. C &\cite{2008Ste01}& PF & Darmstadt & Germany &2008 \\
 & & & & & & & \\
 & & & & & & & \\
$^{165}$Ir & C.N. Davids & Phys. Rev. C &\cite{1997Dav01}& FE & Argonne & USA &1997 \\
$^{166}$Ir & S. Hofmann & Z. Phys. A &\cite{1981Hof01}& FE & Darmstadt & Germany &1981 \\
$^{167}$Ir & S. Hofmann & Z. Phys. A &\cite{1981Hof01}& FE & Darmstadt & Germany &1981 \\
$^{168}$Ir & C. Cabot & Z. Phys. A &\cite{1978Cab01}& FE & Orsay & France &1978 \\
$^{169}$Ir & C. Cabot & Z. Phys. A &\cite{1978Cab01}& FE & Orsay & France &1978 \\
$^{170}$Ir & C. Cabot & Z. Phys. A &\cite{1977Cab01}& FE & Orsay & France &1977 \\
$^{171}$Ir & A. Siivola & Nucl. Phys. A &\cite{1967Sii02}& FE & Berkeley & USA &1967 \\
$^{172}$Ir & A. Siivola & Nucl. Phys. A &\cite{1967Sii02}& FE & Berkeley & USA &1967 \\
$^{173}$Ir & A. Siivola & Nucl. Phys. A &\cite{1967Sii02}& FE & Berkeley & USA &1967 \\
$^{174}$Ir & A. Siivola & Nucl. Phys. A &\cite{1967Sii02}& FE & Berkeley & USA &1967 \\
$^{175}$Ir & A. Siivola & Nucl. Phys. A &\cite{1967Sii02}& FE & Berkeley & USA &1967 \\
$^{176}$Ir & A. Siivola & Nucl. Phys. A &\cite{1967Sii02}& FE & Berkeley & USA &1967 \\
$^{177}$Ir & A. Siivola & Nucl. Phys. A &\cite{1967Sii02}& FE & Berkeley & USA &1967 \\
$^{178}$Ir & A.I. Akhmadzhanov & Bull. Acad. Sci. USSR &\cite{1972Akh01}& FE & Dubna & Russia &1972 \\
$^{179}$Ir & U. Bosch-Wicke & Z. Phys. A &\cite{1992Bos01}& FE & Berlin & Germany &1992 \\
$^{180}$Ir & A.I. Akhmadzhanov & Bull. Acad. Sci. USSR &\cite{1972Akh01}& FE & Dubna & Russia &1972 \\
$^{181}$Ir & A.I. Akhmadzhanov & Bull. Acad. Sci. USSR &\cite{1972Akh01}& FE & Dubna & Russia &1972 \\
$^{182}$Ir & R.M. Diamond & Nucl. Phys. &\cite{1961Dia01}& FE & Berkeley & USA &1961 \\
$^{183}$Ir & A.K. Lavrukhina & Sov. Phys. Dokl. &\cite{1961Lav01}& SP & Dubna & Russia &1961 \\
$^{184}$Ir & V.I. Baranov & Bull. Acad. Sci. USSR &\cite{1960Bar01}& SP & Dubna & Russia &1960 \\
$^{185}$Ir & R.M. Diamond & Nucl. Phys. &\cite{1958Dia01}& LP & Berkeley & USA &1958 \\
$^{186}$Ir & R.M. Diamond & Nucl. Phys. &\cite{1958Dia01}& LP & Berkeley & USA &1958 \\
$^{187}$Ir & R.M. Diamond & Nucl. Phys. &\cite{1958Dia01}& LP & Berkeley & USA &1958 \\
$^{188}$Ir & T.C. Chu & Phys. Rev. &\cite{1950Chu01}& LP & Berkeley & USA &1950 \\
$^{189}$Ir & W.G. Smith & Phys. Rev. &\cite{1955Smi01}& LP & Berkeley & USA &1955 \\
$^{190}$Ir & L.J. Goodman & Phys. Rev. &\cite{1947Goo01}& LP & Ohio State & USA &1947 \\
$^{191}$Ir & B. Venkatesachar & Nature &\cite{1935Ven01}& AS & Bangalore & India &1935 \\
$^{192}$Ir & E. McMillan & Phys. Rev. &\cite{1937McM01}& LP & Berkeley & USA &1937 \\
$^{193}$Ir & B. Venkatesachar & Nature &\cite{1935Ven01}& AS & Bangalore & India &1935 \\
$^{194}$Ir & E. McMillan & Phys. Rev. &\cite{1937McM01}& LP & Berkeley & USA &1937 \\
$^{195}$Ir & D. Christian & Phys. Rev. &\cite{1952Chr01}& PN & Ames & USA &1952 \\
$^{196}$Ir & H. Vonach & Z. Phys. &\cite{1966Von01}& LP & Wien & Austria &1966 \\
$^{197}$Ir & D. Christian & Phys. Rev. &\cite{1952Chr01}& PN & Ames & USA &1952 \\
$^{198}$Ir & A. Szalay & Radiochem. Radioanal. Lett. &\cite{1973Sza01}& LP & Derecen & Hungary &1973 \\
$^{199}$Ir & K. Zhao & Chin. Phys. Lett. &\cite{1993Zha01}& TR & Daresbury & UK &1993 \\
$^{200}$Ir & S.T. Steer & Phys. Rev. &\cite{2008Ste01}& PF & Darmstadt & Germany &2008 \\
$^{201}$Ir & S.T. Steer & Phys. Rev. &\cite{2008Ste01}& PF & Darmstadt & Germany &2008 \\
$^{202}$Ir & S.T. Steer & Phys. Rev. &\cite{2008Ste01}& PF & Darmstadt & Germany &2008 \\
 \\
\end{longtable}

\end{document}